# Interaction of Confined Light with Optically Structured Thin Film Organic Semiconductor Devices

Kuljeet Kaur[1], Pooja Bhatt[1], Ben Johns[1], and Jino George[1*]

[1]Indian Institute of Science Education and Research (IISER) Mohali, Punjab-140306, India.

KEYWORDS: Charge Transport, thin film optics, optical interference, weak and strong coupling, polaritonic states.

**ABSTRACT:** Pioneering experiments of Karl H. Drexhage explained the classical interaction of light with matter and the modification of the decay rates of an emitter.[1] Here, we tried to mimic these experiments in a slightly different configuration and measured the electron mobility of a thin film semiconductor from weak to strong coupling regime. Perylene diimide (organic semiconductor dye) molecules are deposited on a MOSFET device. The refractive index mismatch between the silicon/silicon dioxide layer and the dye molecules forms an interference pattern. The frequency of the interference lines is tuned by changing the thickness of the organic semiconductor. Interestingly, we observed an increase in the electron mobility of the active layer once the system slowly entered into strong coupling condition in a λ cavity. Whereas resonance tuning of a λ/2 cavity does not affect the electron transport, suggesting the system is still in the weak coupling regime. These results are further correlated by optical measurements and transfer matrix simulations. The increase in electron mobility is not large due to high dissipation or low-quality factors of the cavity modes. However, the mirrorless configuration presented here may offer a simpler way of studying the properties of the polaritonic states.

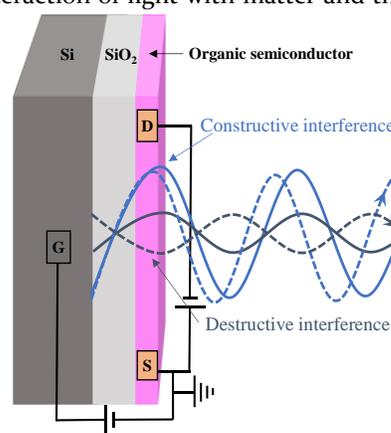

**Introduction:**

Strong light-matter coupling with organic molecules generated a surge of interest among researchers due to its versatile applications.[2] For example, hybrid light-matter states can be generated by coupling an ensemble of small molecules to a cavity mode, thereby affecting their chemical and physical properties.[3] Light-matter interactions are broadly classified as weak and strong interactions.[4] The amount of electromagnetic field experienced by an organic dye molecule purely depends on the environment in which it has been placed, which control the spontaneous emission of the emitter.[5] Surface-enhanced Raman scattering, metal-enhanced fluorescence, etc., are good examples of such near and far field effects.[6] Under weak coupling conditions, molecular energy levels cannot be modified. However, the presence of an oscillating field enhances its decay into the surrounding. Another way to control the light is to structure the medium's refractive index, thereby slowing down the light. Karl H. Drexhage first studied optical interference with a multilayer of dye molecules.[7-8] A monolayer of Eu-complex dye molecules is placed on a multilayer of long-chain fatty acid, and the position is tuned by varying the thickness of the high-index medium. This generates constructive or destructive interference patterns, and the dye molecule feels the electromagnetic field differently as its position changes in the medium. Here, a clear modification in the radiative decay of the emission is observed, and the process is correlated with classical interference of light.[7] This breakthrough experiment considered the classical nature of the light, and the emissivity of the system is purely proportional to the square of the electromagnetic field. Recent studies on electromagnetic field vacuum explain this process much better; a continuum of the field is now becoming discrete, allowing the system to emit directly into these new channels.[9] Can we use this interesting phenomenon to modify the bulk properties of materials?

Strong light-matter interaction offers a way to modify molecules and material properties. The energy exchange between the photon and the electronically excited state of the molecule produces polaritonic states having both the properties of light and matter. In other words, quasi-Bosonic states are formed with a peculiar dispersive nature but have an effective mass much lower than an exciton. Due to these intriguing properties, polaritonic states show Bose-Einstein condensation and polariton lasing, etc., at room temperature.[10-11] Polaritonic states are formed by coupling a large number of molecules to a cavity photon; hence it offers delocalization of molecular excited states.[12-14] This process can enhance the energy/electron transport in the medium.[15-16] This phenomenon is not just limited to electronic energy levels. The recent observations on the modification of exotic behaviour, such as superconductivity and ferroelectricity under vibrational strong coupling (VSC), are evidence of it.[17-18] Another observation is the modification of ionic conductivity of water under VSC of O-H stretching vibrational bands.[19] There are few experimental studies on conductivity enhancement under the electronic strong coupling (ESC) of organic semiconductors.[15, 20-21] There are mainly three mechanisms proposed in the original study suggesting a change in the effective mass, work function variation, and the delocalization of the excited molecular orbitals.[15] A recent study on P-type semiconductors is also proposed to boost transport under ESC.[20] Hence, a clear understanding of the electron transport mechanism is unavailable. The upper polaritonic state (UP) normally decays faster than the lower polaritonic state (LP) and doesn't emit under normal conditions. LP emission can be tailored by controlling the coupling strength of the total system and hence rich in photophysics. For example, intersystem crossing can be effectively enhanced by tuning the cavity mode position or changing the coupling strength.[22-23] Recent observations on the involvement of dark states enrich further applications of polaritonic chemistry.[24-25]

A recent experimental breakthrough suggests that vacuum fluctuations can affect the quantum nature of 2D materials, such as the quantum hall effect.[26] This indicates the modification of electron-electron interaction under strong coupling conditions. Thin film-based organic semiconductors are less ordered, and hopping is the dominating mechanism in such systems. Recent theoretical studies suggest that collective coupling can enhance charge transport efficiently.[12] This suggests orders-of-magnitude enhancement of electron transport through the quasi-Bosonic states and coherence plays a crucial role in controlling the hopping mechanism in the coupled system.[27] Ultra-long propagation of polaritonic states has been demonstrated in a recent study and suggesting that the excitonic diffusion can be ballistic under light-matter strong coupling condition.[28] Similar studies are available with J-aggregates in which the lowest unoccupied molecular orbitals are delocalized up to 10-12 molecules within an aggregate under matter-matter strong coupling.[29] This can be true in the light-matter coupling, but the extent of coherence now depends on the photon (Bosonic nature), and hence a long-range transport can be anticipated. Recent energy transfer studies also underline the above observations.[16, 30] Now, a million-dollar question is what is the strong coupling limit to observe a change in the material properties? Can we modify the bulk transport even under weak coupling conditions?

Conventional metal oxide semiconductor field effect transistor (MOSFET) configuration is used to quantify the electron and hole mobility in organic semiconductors. Most of the time silicon/silicon dioxide (Si/SiO$_2$) layer will be used as the bottom gate. This limits the study of the transmittivity of a Fabry-Perot cavity. There is a recent report on using Si/SiO$_2$ as the substrate by using the advantage of refractive index mismatch between the layers to construct a mirrorless cavity structure. W. L. Barnes and co-workers used this configuration to study the photoswitching of spiropyran to merocyanine.[31] Si/merocyanine/air configuration gives more than 20% splitting energy to enter into the ultrastrong coupling condition. Further such open cavity configuration can clearly modify the absorption envelope of the system, whereas the photoluminescence doesn't follow the polaritonic state line shape.[32] Here, the motivation is to use a mirrorless cavity configuration as a test bed for studying the conductivity of a strongly coupled system. The major advantage of this configuration is that the top mirror is absent; hence, the 3-terminal MOSFET config-uration can be used directly to extract the electron mobility of the active layer. Cavity tuning can be done by varying the thickness of the active layer. Here, Si/SiO$_2$/PDI/Air confine the electromagnetic field and couple the absorption band of the active PDI molecules.

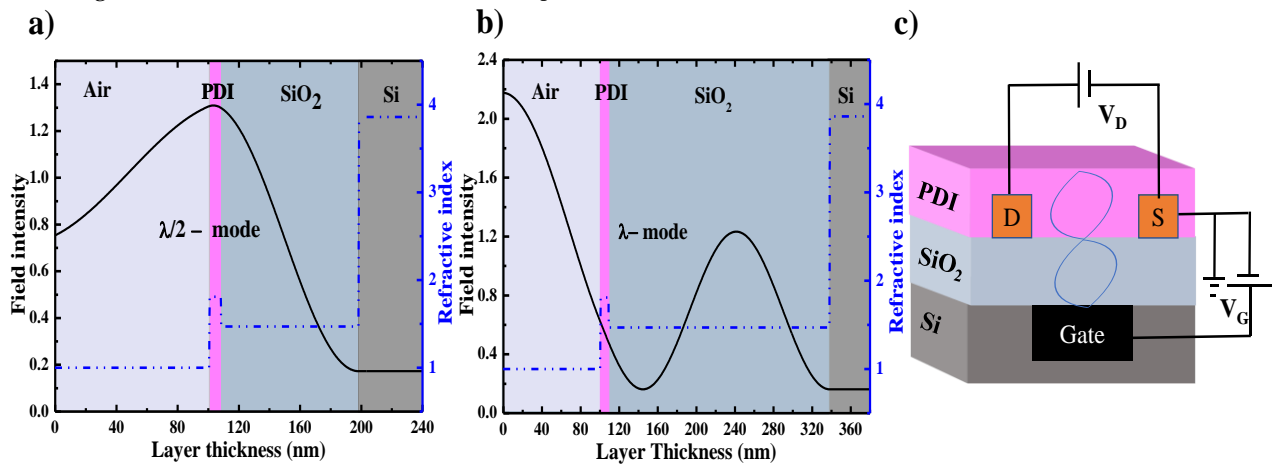

**Figure 1. Mirrorless MOSFET- Fabry-Perot cavity**: Electric field distribution profile with refractive index variation of different layers in Air/PDI (8 nm)/SiO2/Si configuration at 565 nm for (a) λ/2 mode cavity (90 nm SiO2) and (b) λ-mode cavity (230 nm SiO2). (c) Schematic of MOSFET device, with different layers in FP-configuration without mirrors (Active layer (PDI) is represented in pink colour).

Earlier studies used PDI thin films spin-coated on plasmonic arrays for electron transport measurements.[15] The separation between the source and drain is kept above 50 μm to eliminate the parasitic contact resistance completely. This will help to read the actual conductivity of the active layer. Tuning experiment with a varying periodicity of the plasmonic arrays clearly shows an ON-resonance effect and almost an order of magnitude increase in the electron mobility under ESC. Solution-processed PDI thin film typically shows a gate mobility of 10$^{-3}$ cm$^2$/V/sec in a bottom gate MOSFET device.[33] Changing the thickness of the PDI by varying the concentration can affect the mobility due to molecular packing. Mobility can be further enhanced by annealing the thin film at elevated temperatures. Molecular packing has a good role in conductivity enhancement in organic thin films, and possibly the polaritonic states formed under ESC of the PDI molecules can also boost the transport in the system. Very recently, we tested this argument using highly crystalline 2D materials and found that the electron mobility is even enhanced more than 50 times under resonance conditions.[34] These experiments are done in a half-mirror cavity configuration. The high absorption coefficient and easy solution processability make PDI molecules the best candidate to test this idea. Here, we clearly observed a change in the electron mobility in mirrorless MOSFET-FP (Fabry-Perot) cavities while coupling the electronic state of PDI semiconductors to a confined mode.

## Results and Discussion

Mirrorless cavities are prepared by spin-coating PDI solution on a MOSFET and annealing the sample at ambient conditions. PDI molecules are synthesized using a standard procedure reported in the literature.[15] The dicyanated PDI is recrystallized multiple times and further characterized by NMR measurements (section 2 of the SI). Spin coating is done using chloroform as a solvent (PDI concentration varies between 0.05 to 1.0 wt%), and the desired thickness is achieved by coating them at 750 rpm in 60 seconds using a tabletop spin coater (experimental methods). Due to the high refractive index (n= 1.82) of the active layer (PDI), the light gets confined to form leaky optical modes between the silicon and the air interface. There are mainly two thicknesses of SiO$_2$ used for our studies that can support $\lambda/2$ (90 nm) and $\lambda$ cavity (230 nm; see figure 1). These three terminal devices are used for all optical and electrical measurements (figure 1c). 90 nm SiO$_2$ device shows a shallow interference mode with an electric field maximum at the active layer position for an 8 nm PDI thin film (figure 1a and 2a). Reflectance spectra acquired show both the features of PDI molecules and the shallow cavity modes. 230 nm SiO$_2$ device shows a clean interference pattern for the same PDI thickness. The electric field maximum is within the SiO$_2$ medium and moves slowly towards the PDI layer upon increasing the thickness of the active layer (figure 1b). PDI thickness variation result in a gradual change in the electron mobility as the crystallinity of the PDI is better in the lower concentrations.[33]

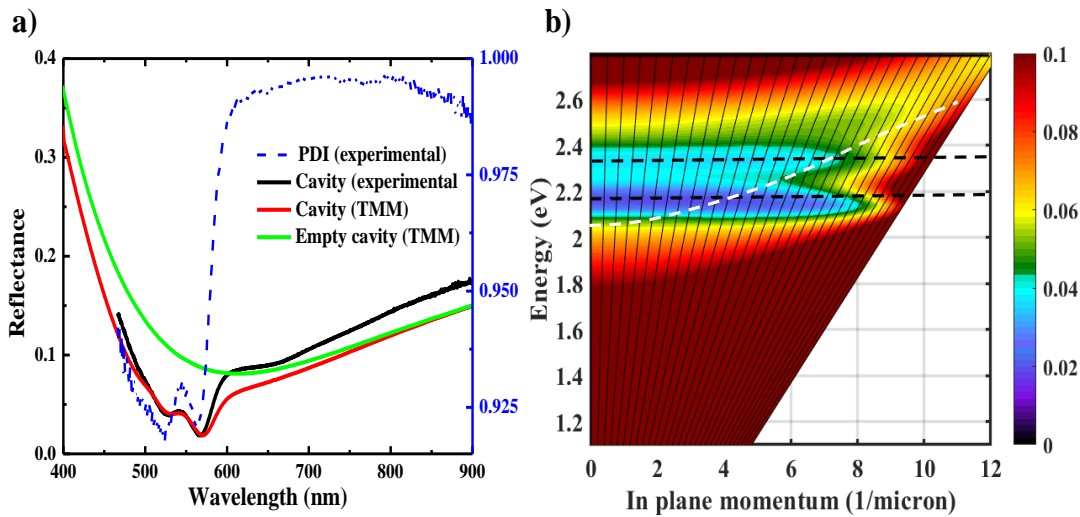

**Figure 2. λ/2-Cavity (Optical output):** (a) Reflectance spectra of PDI (blue colour), experimental spectra of the coupled system (black colour), and TMM fitting with PDI (8 nm) (red colour) and without PDI absorption (green colour). (b) The simulated dispersion plot shows the broadening of the modes at a higher angle. Dotted lines indicate the exciton frequency (black colour) and light line of the empty cavity mode (white colour).

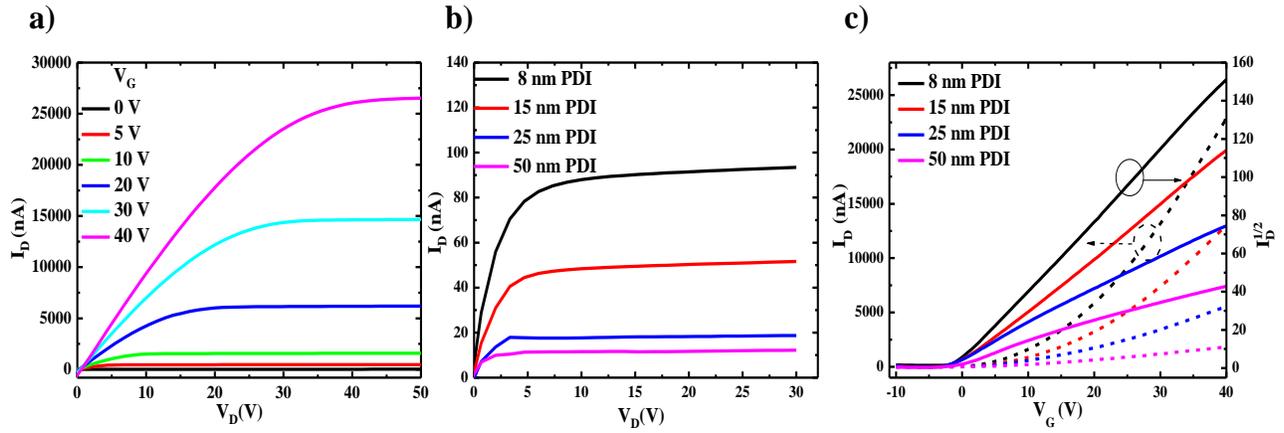

**Figure 3. λ/2-Cavity (Electrical output):** (a) I-V-characterization of 8 nm thick PDI on MOSFET-FP mirrorless cavity (90 nm SiO2) at different gate voltages. (b) Output characteristics for different thicknesses of PDI at zero gating voltage, and (c) The corresponding transfer characteristics of the system at $V_D$=40 V.

The leaky modes formed in the λ/2 cavities show a very broad signature and cross the PDI absorption region for the thinnest layers. The molecular concentration is very low here to achieve strong coupling conditions. PDI thin film shows two absorption maxima at 525 and 565 nm with a full-width half-maximum of 0.45 eV (figure 2a). Increasing the thickness of the PDI moves the resonance position to a higher wavelength, and hence the chance of getting a resonance situation with λ/2 is not possible in this condition. This has been further confirmed by transfer matrix simulation (TMM) dispersion and phase intensity correlations (figure 2b and section 5 of SI). The absorption linewidth is little affected in the beginning and starts showing the formation of weak polariton branches for very thin layers of PDI. Phase calculation also supports this argument that as the thickness increases, the cavity mode moves away from the resonance position (figure S7). TMM simulation of the λ/2 cavities show the modes are very leaky with a Q-factor of 1.53; hence, the chance of strong coupling is feeble at such conditions.

Electrical measurements are conducted further to understand the effect of weak interference on the bulk material properties. I-V measurements varying the gate voltage from 0 to 40 V give field effect characteristics, with the drain voltage reaching saturation close to 30 V (figure 3a). A stock solution of PDI (0.5 wt%) is prepared in chloroform and further diluted to five different concentrations from the same stock each time. Thin films fabricated on 90 nm SiO$_2$ devices are further measured using a probe station kept in an inert atmosphere (O$_2$ <10 ppm; H$_2$O <10 ppm). Comparison of the I-V measurements suggests that the conductivity of the thinnest layer is maximum and decreases gradually, and becomes constant for a neat layer of 0.5 wt% sample. Both output and transfer characteristics suggest that the electron mobility is high for the thinnest layer (figure 3b, c). Please note that the electron mobility is extracted at the saturation regime (drain voltage $V_D$= 40 V). Further details on the mobility calculation and the output/transfer characteristic are given in section 7 of the SI. At least half a dozen repeated measurements are done with freshly casted films for all the thicknesses, and the mobility data is averaged (section 8; table T1 of the SI). Kindly note that more than 100 independent devices are measured to reach a general conclusion. All the raw optical/electrical data, along with the standard deviations, are given in the SI. Average mobility versus empty mode position (calculated using TMM; SI) is given in figure 6a for further comparison.

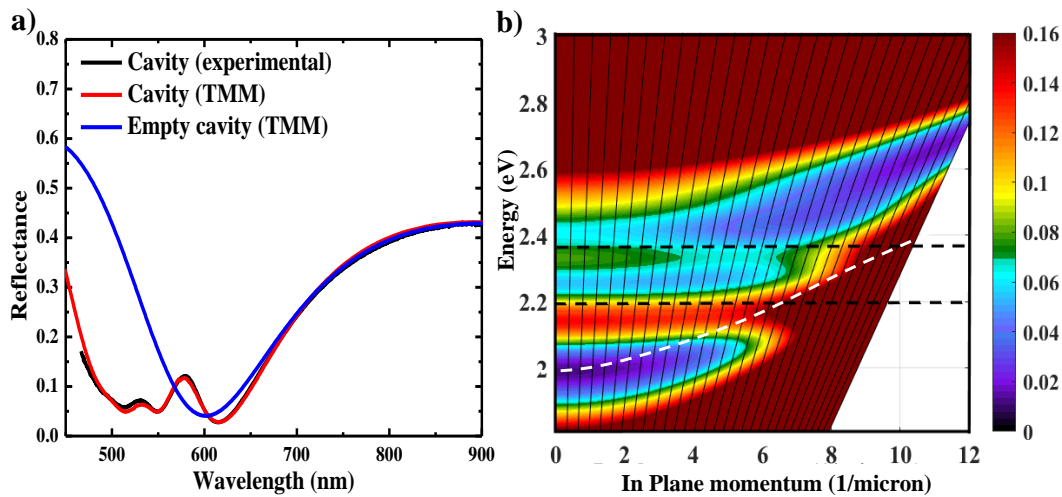

**Figure 4. λ-Cavity (Optical output):** (a) Reflectance spectra of PDI (black colour), and TMM fitting with PDI (55 nm) (red colour) and without PDI absorption (blue colour). (b) The simulated dispersion plot shows the formation of polaritonic states. Dotted lines indicate the exciton frequency (black colour) and light line of the empty cavity mode (white colour).

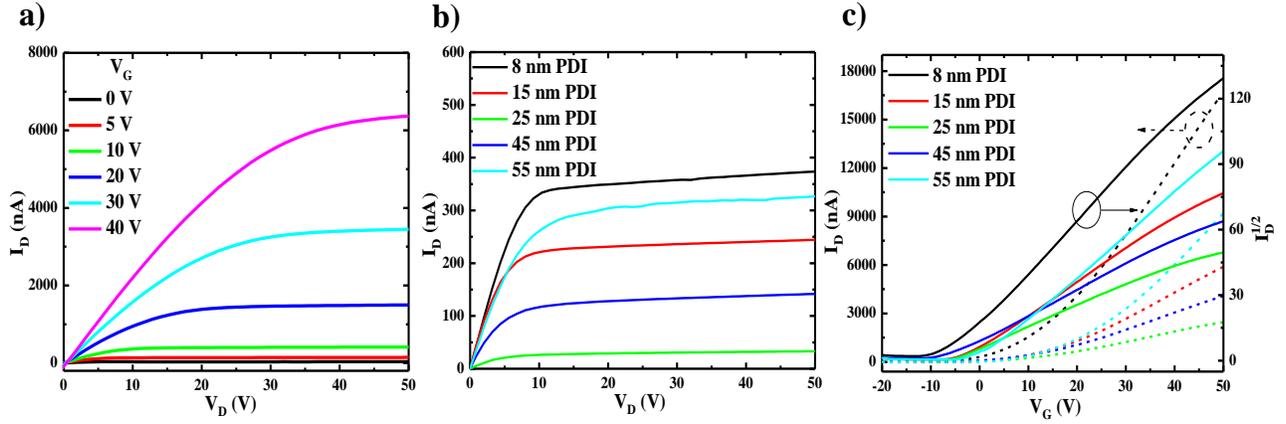

**Figure 5. λ-Cavity (Electrical output):** (a) I-V-characterization of 55 nm thick PDI on MOSFET-FP mirrorless cavity (230 nm SiO2) at different gate voltages. (b) Output characteristics for different thicknesses of PDI at zero gating voltage, and (c) The corresponding transfer characteristics of the system at $V_D$=40 V.

The above experiments show that electron mobility is not influenced due to the presence of leaky λ/2 cavity modes, but set as a control to test the same experiment in the λ cavity. Here, the main modification of the MOSFET device is that the SiO₂ layer thickness is 230 nm and the PDI thickness varies from 8 to 65 nm. λ cavities show better-confined modes with a Q-factor of 3.45, as shown in figure 4a. The cavity linewidth and the natural linewidth of the molecule are comparable in this condition. An increase in the PDI thickness causes a shift in the λ cavity mode position and reaches a resonance situation at 35 nm of the active layer. However, the absorption strength is lower at this condition; the polaritonic branches don't evolve till the PDI thickness increases to 35 nm (figure S9). A clear signature of polariton dispersion is observed at 55 nm of PDI, indicating the system is entering into strong coupling regime (figure 4a). Please note that the PDI absorption band is broad enough to couple the cavity mode at 55 nm PDI thickness, and the empty mode is now close to 590 nm, as obtained from the TMM calculation (figure S10). The above experiments clearly show that the system is slowly moving from weak to strong coupling regime around this PDI thickness, and a further increase in the active layer thickness will move the cavity mode away from the molecular resonance. We proved the interplay of the collective strength of the active layer and the formation of polaritonic branches by analyzing the dispersion, Hopfield coefficients, and phase correlation of the system with and without strong coupling (sections 5 and 6 of the SI).

In order to understand the effect of strong coupling, we tested the electrical output of the λ cavity. A very similar trend is observed for the thinner PDI layers till 35 nm, and further increasing the thickness increases the conductivity of the system (figure 5). Here, mobility measurements are done on half-a-dozen of samples and averaged for better comparison (sections 7 and 8 of SI). Electron mobility extracted in the saturation regime shows that 55 nm PDI shows a second maximum. This very interesting observation is further correlated with the optical measurements in figure 4. PDI active layer of 35 nm entered into the minimum criteria for strong coupling (g=0.17 eV > ($\Gamma_c$ - $\Gamma_m$)/2), where g is the interaction term, $\Gamma_c$ is the cavity FWHM (0.66 eV), and $\Gamma_m$ is the molecular FWHM (0.45 eV).[4] Here, the branching begins with two distinct polaritonic branches at 55 nm with equal mixing of photon and exciton fractions at a higher angle (figure 4 and S11). The signature of strong coupling is evident from the electrical measurements, and upon increasing the active layer thickness, the electron mobility decreases further, as expected from the optical studies. The electron mobility extracted from the gating experiments is roughly three times, which may be due to leaky cavity modes compared to the mirrored FP cavities and other plasmonic structures. However, modulating the electron mobility just by varying the cavity resonance (PDI thickness) is an interesting observation. This suggests that tailoring of a bulk property can be achieved in a simple MOSFET configuration without introducing any mirrors. Parasitic resistance, electrode leaking, etc., can be avoided here, giving a better field effect from the active layer under strong coupling.

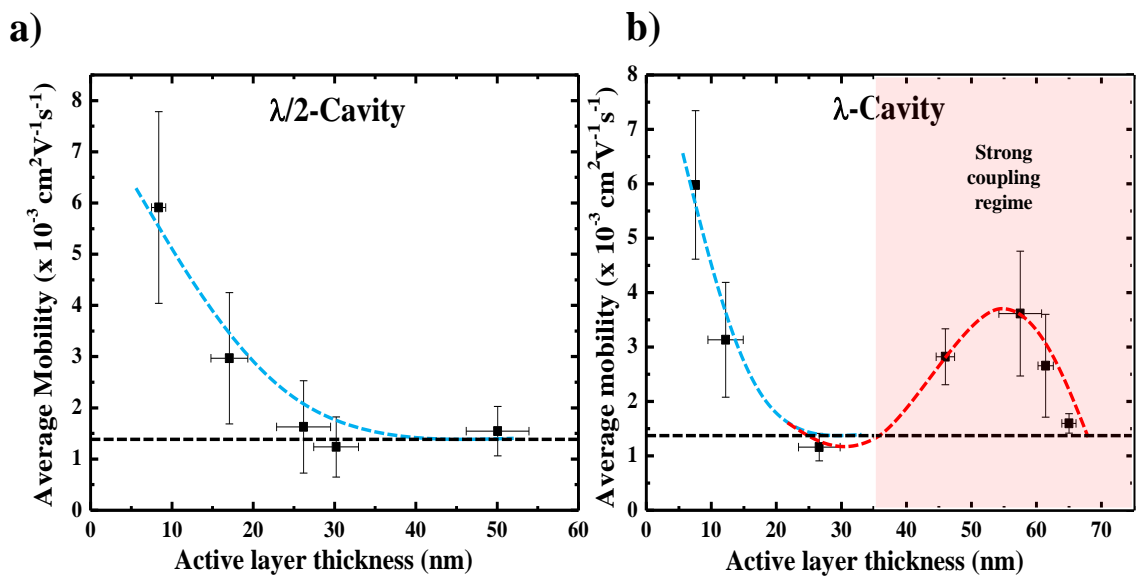

**Figure 6. Electron mobility comparison**: (a) For λ/2-cavity and (b) λ-cavity MOSFET devices. Electron mobility is averaged for multiple measurements on different samples with PDI thin films of thicknesses between 8 nm to 65 nm. Blue, red dotted lines are guides to the eye, and the black dotted line is the minimum gate mobility of the uncoupled device.

Here, in a series of measurements, we varied the thickness of the active layer (PDI) and compared the optical interference generated in the MOSFET structure. Both the optical and electrical measurements suggest that the system must enter into strong coupling regime to see a physical property change (in this case, electron mobility) of the coupled system. λ/2 cavity structure forms a very shallow mode and even if an ON-resonance situation cannot bring a branching of polaritonic states in the system (Figure S6). Increasing the thickness of the PDI moves the resonance position away from the molecular absorption envelope. Field distribution curve as well as the dispersion measurements equally support this argument. Kindly note that the lowest thickness of the sample gives maximum electron mobility, gradually becoming a constant for the larger thickness of the PDI molecules due to the better morphology and packing of the thin films at lower concentrations. Bottom gate mobility extracted for similar samples also show the same behaviour.[33] λ cavity structures also show a similar trend suggesting the morphological characteristic of the PDI (blue dotted line in Figure 6). Grazing incidence wide-angle X-ray scattering experiments and AFM measurements suggest a morphology change of the PDI upon varying their concentration in chloroform (Figures S3 and S4). Further, the electron mobility becomes a constant (1.4 x 10$^{-3}$ cm$^2$/V/sec) at a higher sample concentration (black dotted line). In the case of λ cavity structures, PDI thickness of 35 nm gives an ON-resonance situation (red dotted line; Figure S8-S10). However, the system slowly begins to branch at this condition, and at higher thickness, the PDI molecules are enough in numbers to achieve strong coupling condition. Electron mobility shoots slowly and becomes 3.7 x 10$^{-3}$ cm$^2$/V/sec (average of 17 independent devices) at 55 nm of PDI. This condition is the best situation where the polaritonic states control electron transport. Further increasing the thickness of PDI move the mode away from the absorption envelope and goes OFF-resonance for all higher thickness of the active layer.

## Conclusions

Finally, getting motivated by the experiments of Karl H. Drexhage, we conducted the optical and electrical measurements of an organic semiconductor in an optically structured medium. Optical interference formed in this configuration is well characterized both experimentally and using TMM simulation. The system can only enter into a strong coupling regime for a λ cavity configuration due to the poor absorption strength of the active layer and leaky cavity mode. Here, a commercially available MOSFET device is used to test the idea. Interestingly, the strong coupling regime shows an enhancement in the electron mobility supporting the original finding of vacuum engineering of materials. Dark current measurements done on these mirrorless geometries unequivocally prove the existence of vacuum fluctuations and open new avenues for direct application in polaritonic state-based optoelectronic devices.

## Experimental methods

**Device preparation:** Mirrorless cavities were fabricated by spin coating the desired thickness of PDI dye on bottom gate bottom contact highly doped silicon substrates (purchased from Fraunhofer-IPMS, Germany) having silicon dioxide dielectric layer (90 or 230 nm thickness) and pre-patterned Ti/Au electrodes. Different concentrations of PDI solutions were prepared by dissolving PDI in chloroform and then filtered by using PTFE filters (0.45 μm diameter). n$^{++}$-Si/SiO$_2$-substrates were washed with electronic-grade acetone and isopropyl alcohol and then blow-dried using a nitrogen gun. Then each solution was spin-coated (LABSPIN 6, Suss-Microtech) for 60 seconds at 750 rpm on these substrates to achieve the desired thickness. Samples were post-annealed at 100 °C for 30 min in ambient conditions for better alignment of the molecules.

Reflectance measurements were done using Nikon inverted microscope (Eclipse Ti2) at room temperature using a halogen lamp as a light source. The light reflected by the substrate was collected through a 20x objective lens (0.45 NA) to a spectrometer (SpectraPro HRS-300, Princeton Instruments), which is coupled to a charge-coupled device (CCD) camera (Pylon 100BX) (cooled to -120°C using liquid nitrogen) for spectral acquisition.

FET devices (three-way, bottom gate bottom contact) were electrically characterized inside the glove box under inert conditions (O2 and H2O level less than 10 ppm) using MB 150, Cascade Microtech probe station and Keithley 2636B instrument controlled by kickstart488 software. Output characteristics ($I_D$-$V_D$) were extracted by varying gate voltage from 0 V to 50 V, and transfer characteristics ($I_D$-$V_G$) were traced in a saturation regime at 40 V of drain voltage. Field effect mobility was calculated in the saturation regime using the following formula;

$$\mu = \frac{d\sqrt{I_D}}{dV_G} \frac{2L}{W C_i}$$

where, L (Length of the gold electrodes) = 20 μm, W (width of the electrodes) = 2 mm, $C_i$ (capacitance of gate dielectric) = 3.84 x $10^{-8}$ Fcm$^{-2}$ (for 90 nm SiO$_2$) and 1.5 x $10^{-8}$ F cm$^{-2}$ (for 230 nm SiO$_2$).

## ASSOCIATED CONTENT

Synthesis and characterization of the PDI molecules, morphological studies, TMM simulatuions and all the raw data of optical and electrical measurements (λ/2 and λ cavity) are given in the supporting information.

### AUTHOR INFORMATION

**Corresponding Author**

* jgeorge@iisermohali.ac.in

**Funding Sources**

MoE-Scheme for Transformational and Advanced Research in Sciences (**MoE-STARS/STARS-1/ 175**).

### ACKNOWLEDGMENT

The authors thank Dr. Thibault Chervy for help in the TMM calculations. K. K., P.B., and B.J. thank IISER Mohali for the fellowship. J. G. Thank IISER Mohali for the start-up grant.

# Supporting Information

# Interaction of Confined Light with Optically Structured Thin Film Organic Semiconductor Devices

Kuljeet Kaur[1], Pooja Bhatt[1], Ben Johns[1], and Jino George[1]*

[1]Indian Institute of Science Education and Research (IISER) Mohali, Punjab-140306, India.
*jgeorge@iisermohali.ac.in

**Table of contents**

# Section-1

**Synthesis of perylene-di-imide molecules**: PDI2EH-CN$_2$ dye was synthesized from PDI2EH-Br$_2$ derivative and Cu(I)CN (Figure S1).[1] 2.589 mmol (2 g) of PDI2EH-Br$_2$ and 47.141 mmol (4.22 g) of Cu(I)CN were taken in 250 mL round bottom flask, followed by the addition of 102 mL of dry DMF and heated to 150$^0$ C for 8 hrs under nitrogen atmosphere. The crude product was then concentrated using a vacuum pump and chromatographed on silica using chloroform/ethyl acetate. The final product was then recrystallized multiple times from chloroform with methanol.

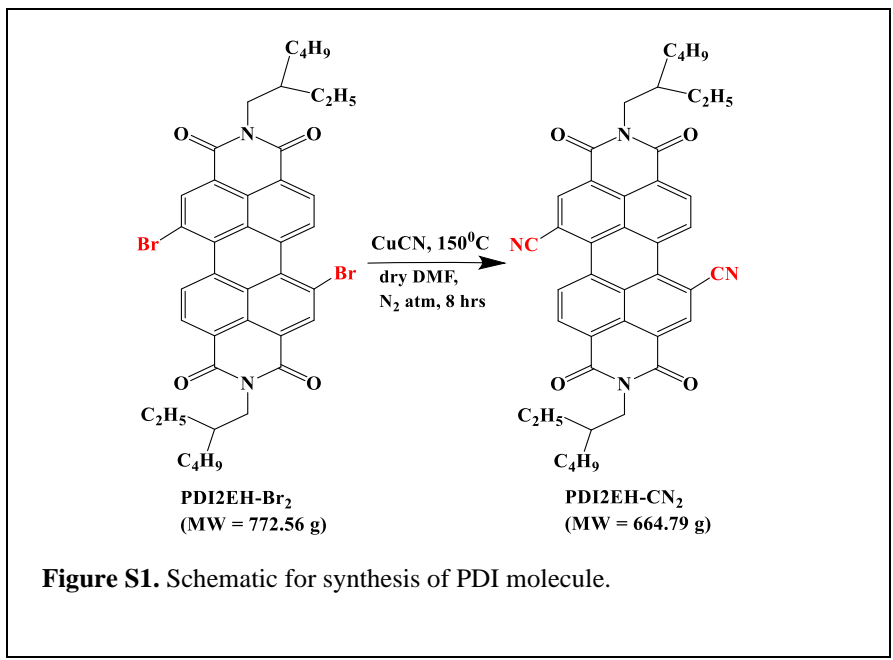

**Figure S1.** Schematic for synthesis of PDI molecule.

# Section-2

**Characterization and optical studies:** $^1$H NMR-spectra of the final product in CDCl$_3$ showed the following peaks, confirming the required product's formation.

1H NMR (CDCl3): δ 9.69-9.65 (d, 2H), 8.95 (s, 2H), 8.92- 8.88 (d, 2H), 4.22-4.16 (t, 4H), 1.89-0.89 (m, 30H).

Absorption spectra of a thin solid film of PDI coated on optically clean glass substrates show two peaks at 565 nm and 525 nm corresponding to 0-1 and 1-1 transitions, respectively (figure S2 (a)). Corresponding reflectance spectra were obtained for various thin films of PDI by spin-coating on Si-substrates (having 90 nm and 230 nm SiO$_2$) at ambient temperature (figure S2 (b) and (c)).

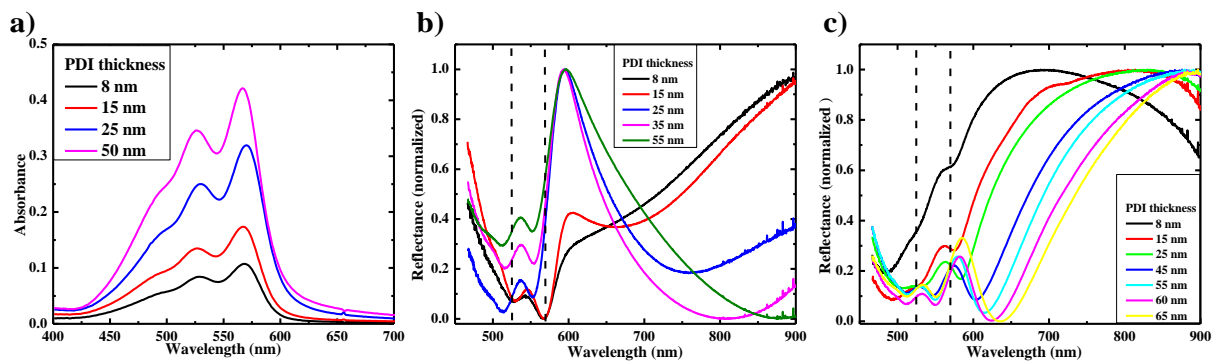

**Figure S2. Optical studies (experimental) on different substrates:** (a) Absorption of thin films coated on optically clean glass substrates. (b) and (c) reflectance spectra of PDI thin films coated on 90 nm and 230 nm $SiO_2$ to study $\lambda/2$ and $\lambda$-cavity, respectively.

## Section-3

**XRD-data:** Grazing incidence wide-angle X-Ray scattering (GiWAXS) measurements were done by keeping the sample to a detector distance of 92.1 nm using an incident wavelength of 0.154 nm (figure S3). PDI solution was spin-coated on 90 nm $SiO_2$-coated silicon substrates (without electrodes) for this study.

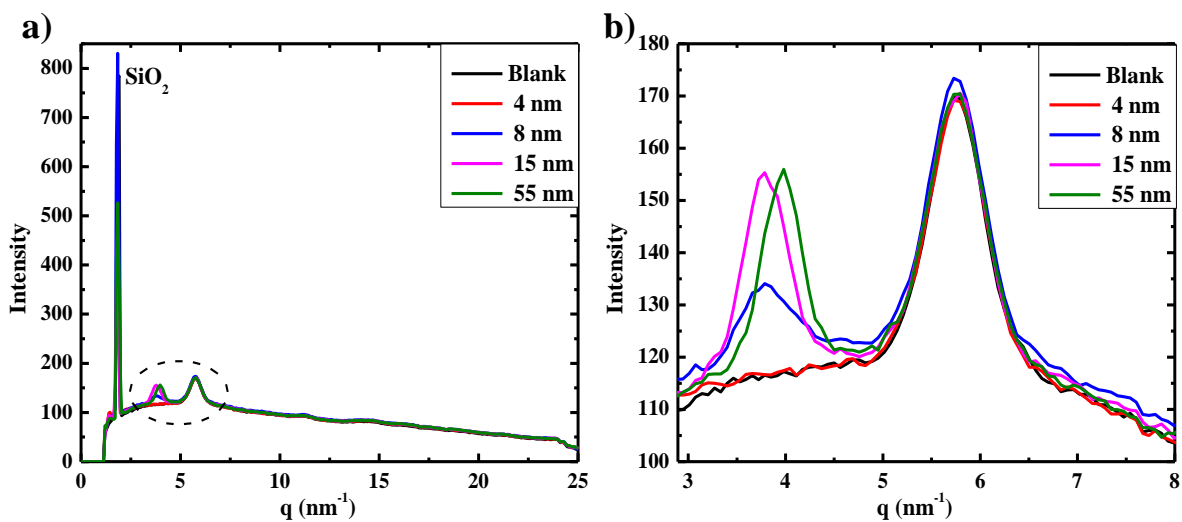

**Figure S3. XRD-studies of thin films of different thicknesses of PDI:** (a) GiWAXS-pattern of PDI films and (b) zoomed image of dotted area of figure (a).

## Section-4

**Atomic force microscope (AFM) data:** AFM studies of spin-coated thin films on 90 nm $SiO_2$-coated Si-substrates were done to check the film morphology (figure S4), which suggested better morphology in thinner PDI films.

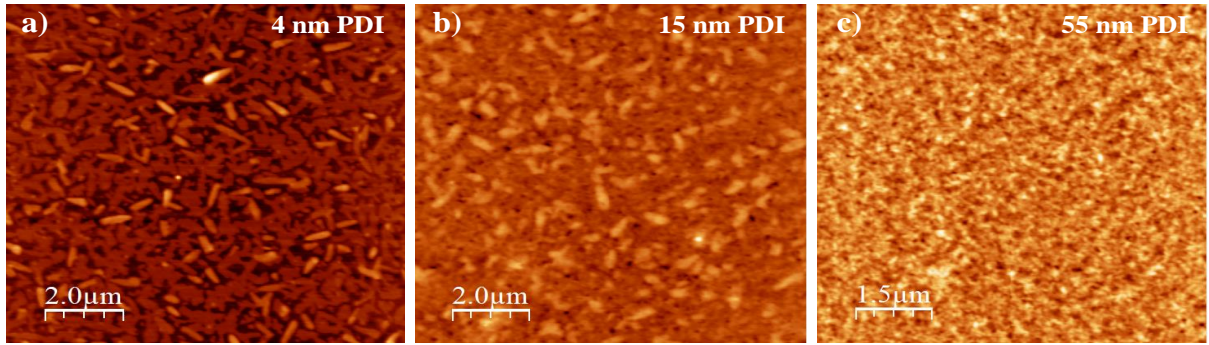

**Figure S4:** AFM of PDI thin films of thickness (a) 4 nm (b) 15 nm (c) 55 nm.

# Section-5

**Transfer matrix method simulation:** The transfer matrix method (TMM) tool is used to study the propagation of electromagnetic waves through multilayer systems. The difference in refractive indices between the layers (Air/PDI/SiO$_2$/Si) leads to the formation of leaky modes. Thicknesses of PDI thin films are estimated by TMM fitting of experimental reflectance spectra of each sample ($n$SiO$_2$=1.46; $n$Si=3.95 at 589 nm). Further, dispersion and field distribution plots were also extracted from TMM for both λ/2 and λ-cavities.

(i) <u>Spectra for λ/2 cavity</u>

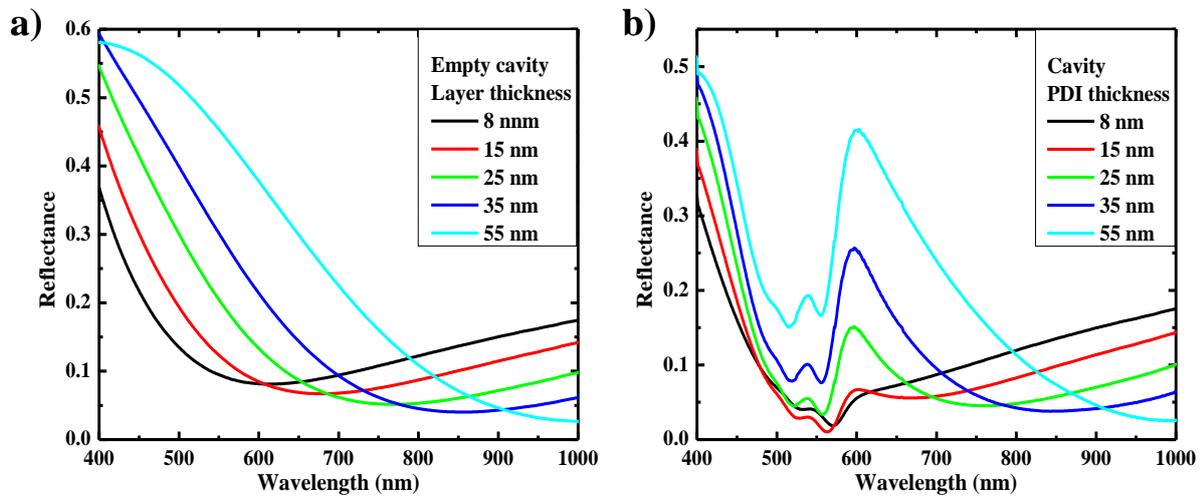

**Figure S5:** a) Reflectance in absence of absorber and b) in presence of absorber (PDI dye) for different thicknesses.

(ii) <u>Dispersion for λ/2 cavity</u>

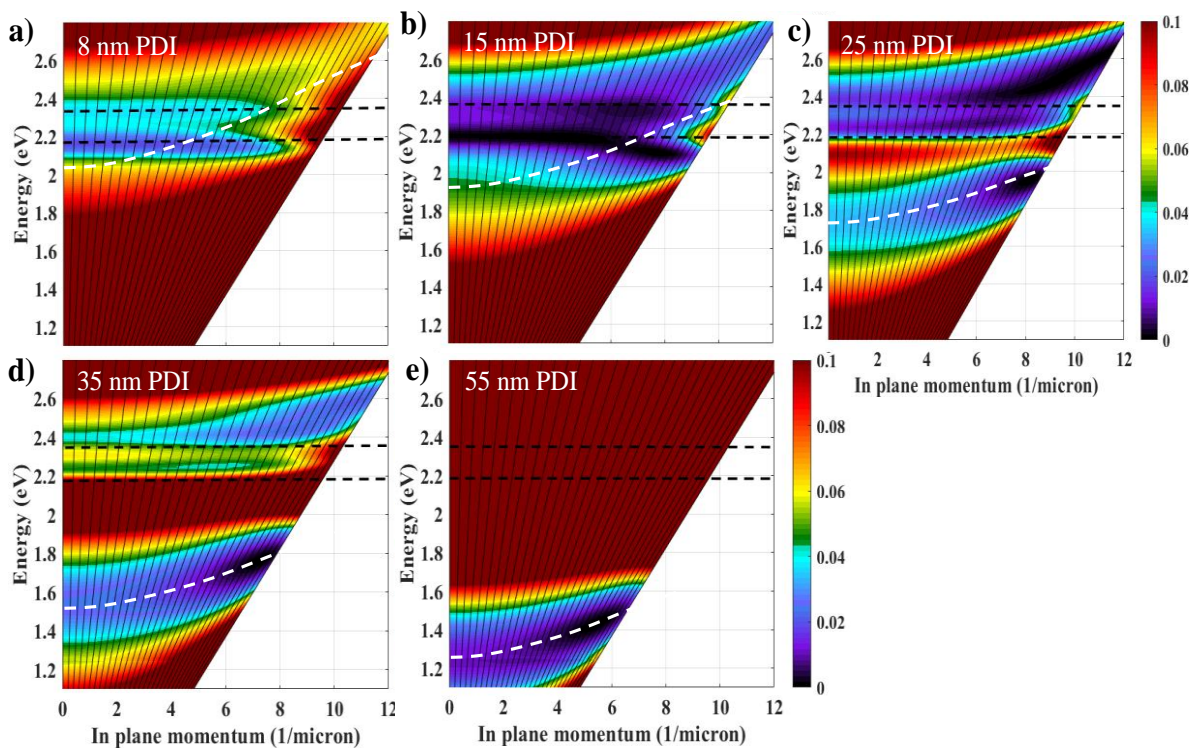

**Figure S6.** Dispersion plots for different thicknesses of PDI molecules for λ/2-cavity. Black dotted lines represent exciton positions and white dotted lines represent empty cavity dispersion. Z-axis represents reflectance.

### (iii) Field intensty distribution plots for λ/2 cavity:

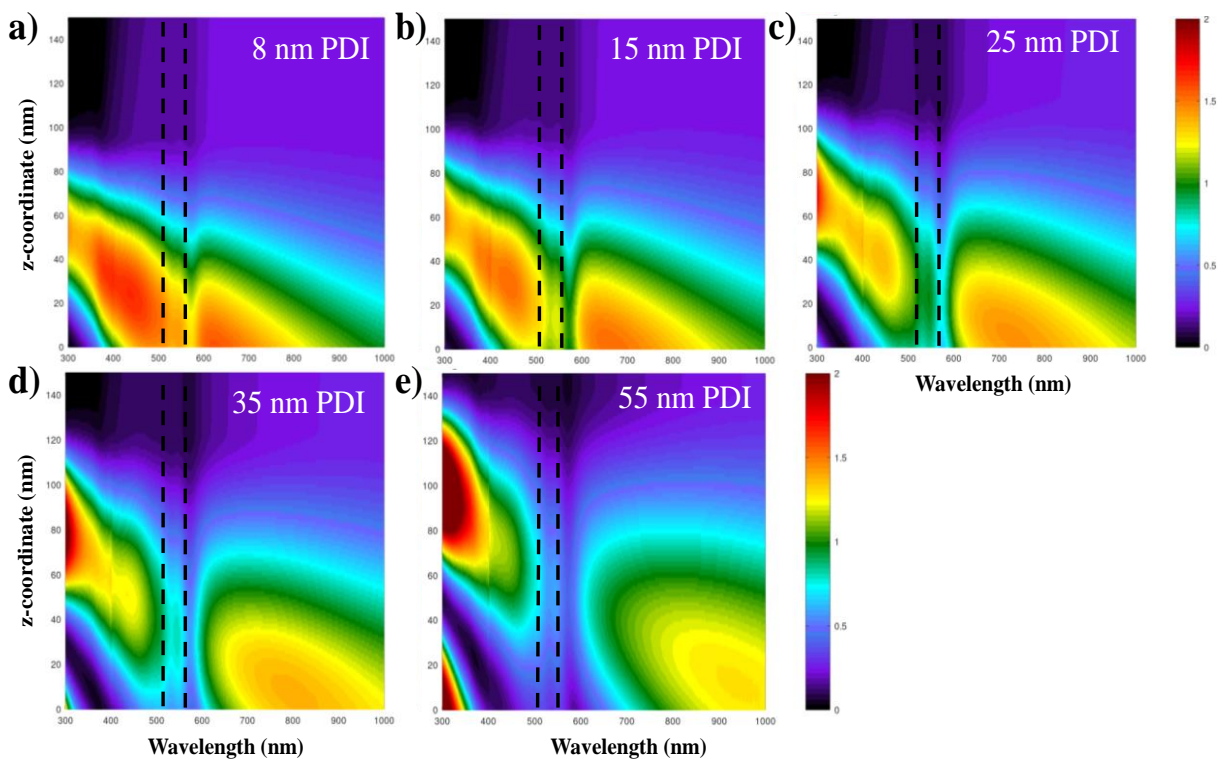

**Figure S7:** Field intensity distribution at different thicknesses for λ/2-cavity with PDI molecules. z-axis represents reflectance and black dotted lines represents exciton positions.

### (iv) **Spectra for λ-cavity**

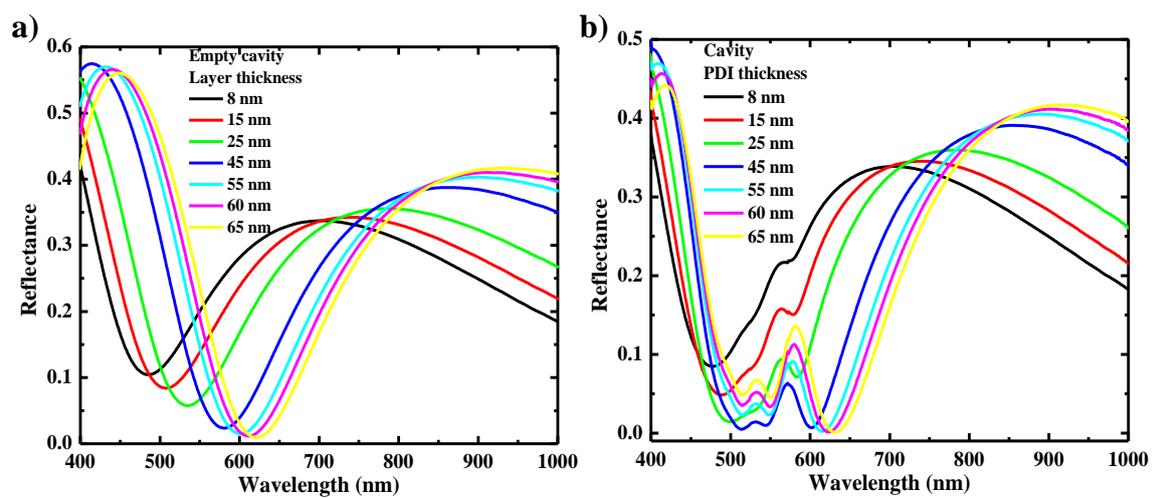

**Figure S8.** (a) Reflectance in absence of absorber and (b) in presence of absorber (PDI dye) for different thicknesses.

**(v) Dispersion for λ cavity**

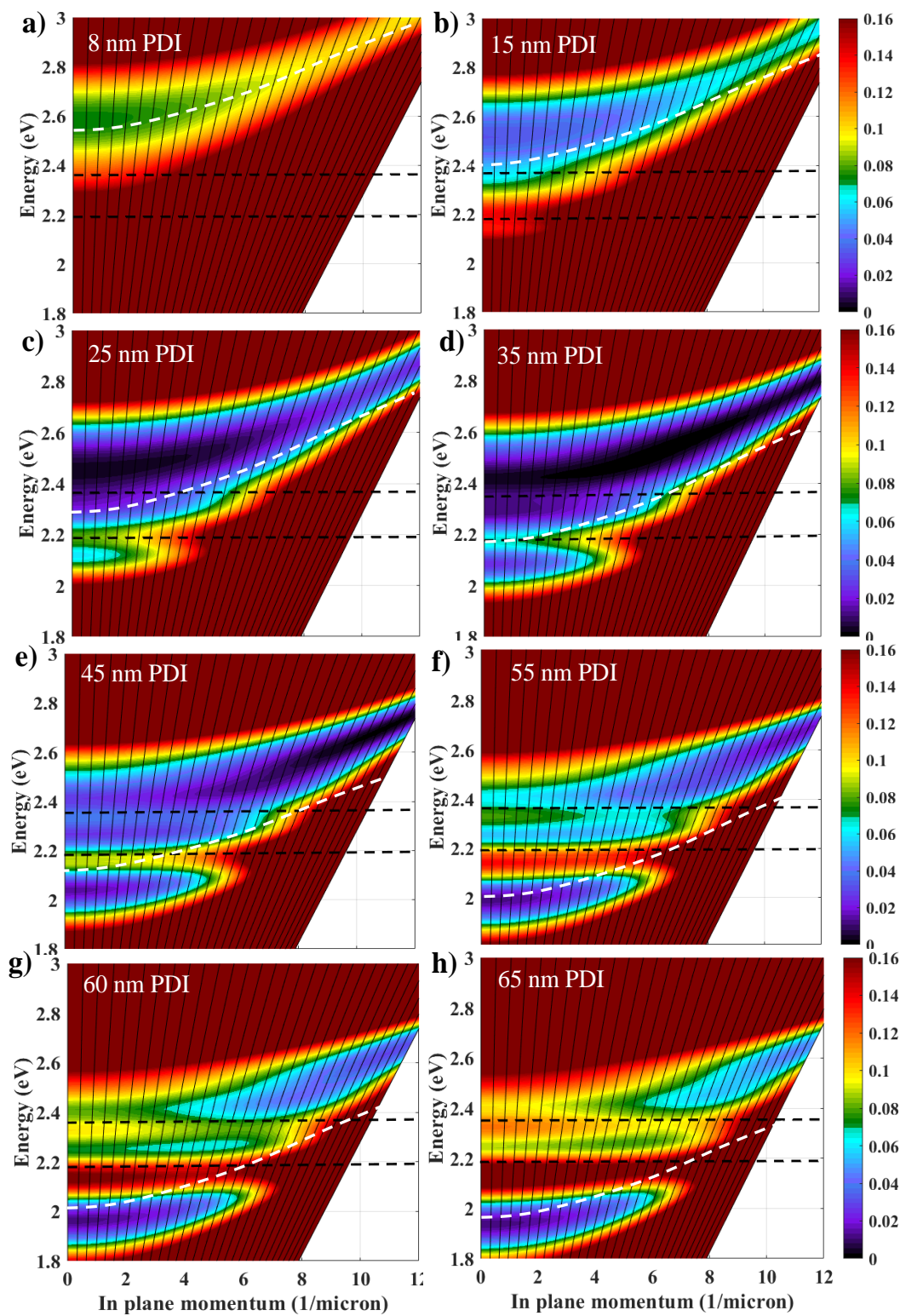

**Figure S9:** Dispersion plots for different thicknesses of PDI molecules for λ-cavity. Black dotted lines represent exciton positions and white dotted lines represents empty cavity dispersion. z-axis represents reflectance.

**(vi)** <u>**Field intensity distribution plots for λ cavity:**</u>

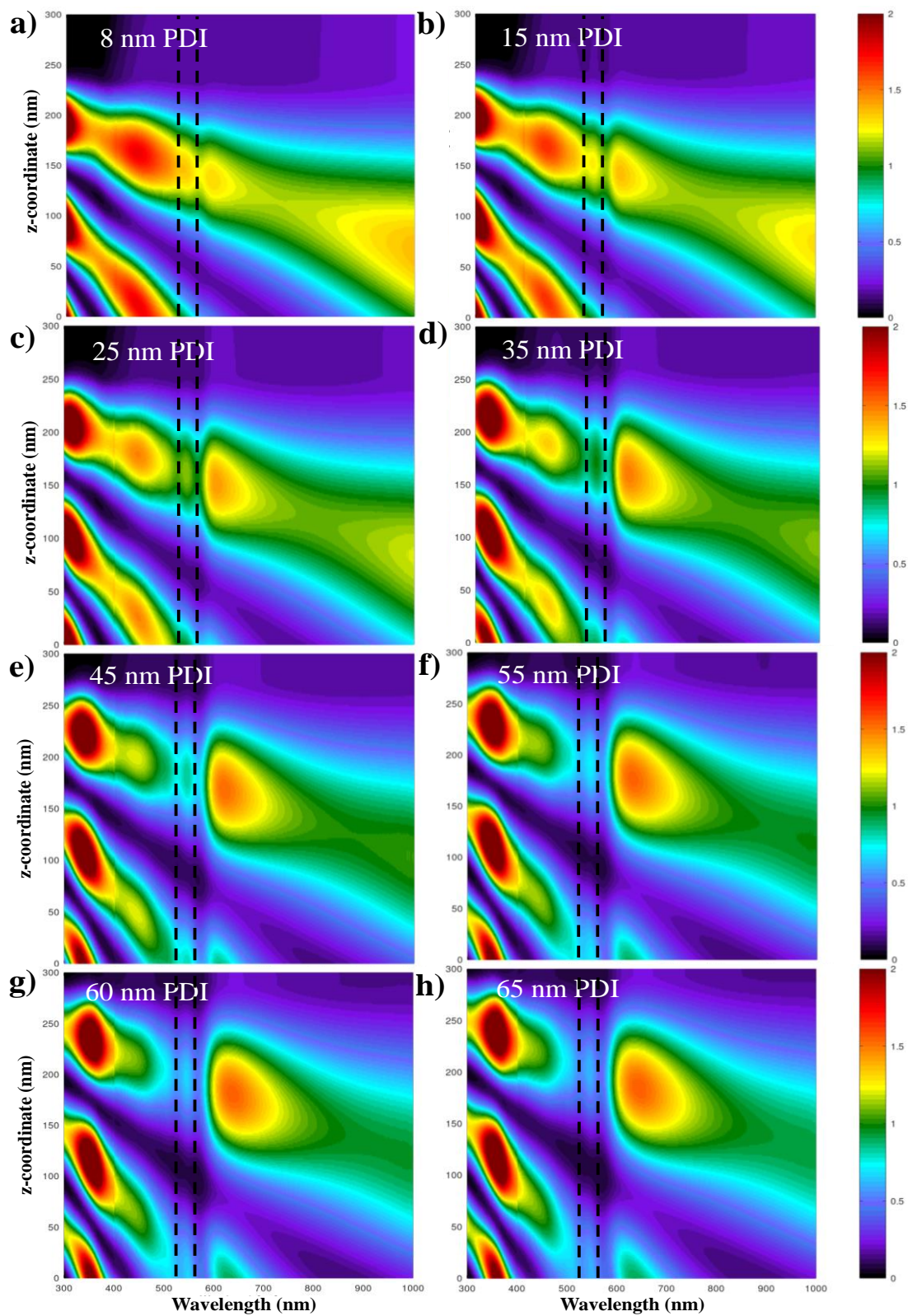

**Figure S10.** Field intensity distribution at different thicknesses for λ-cavity with PDI dye. z-axis represents reflectance and black dotted lines represents exciton positions for PDI dye.

## Section-6

**Hopfield coefficient calculation:** Hopfield coefficient is calculated for 8 nm PDI for λ/2 cavity (Figure S11 (a)) and 35 nm and 55 nm PDI for λ-cavity (Figure S11 (b, c)) by using the following equations[2]:

$$|C|^2 = \frac{1}{2}\left[1 - \frac{(E_{ph}-E_{ex})}{\sqrt{(E_{ph}-E_{ex})^2+(\Omega_R)^2}}\right]$$

$$|X|^2 = \frac{1}{2}\left[1 + \frac{(E_{ph}-E_{ex})}{\sqrt{(E_{ph}-E_{ex})^2+(\Omega_R)^2}}\right]$$

Here, $|C|^2$ represents photonic content, and $|X|^2$ represents excitonic content. $E_{ph}$ and $E_{ex}$ are photons and exciton energies, respectively, and $\Omega_R$ is splitting energy. $E_{ex}$ is taken as 2.275 eV.

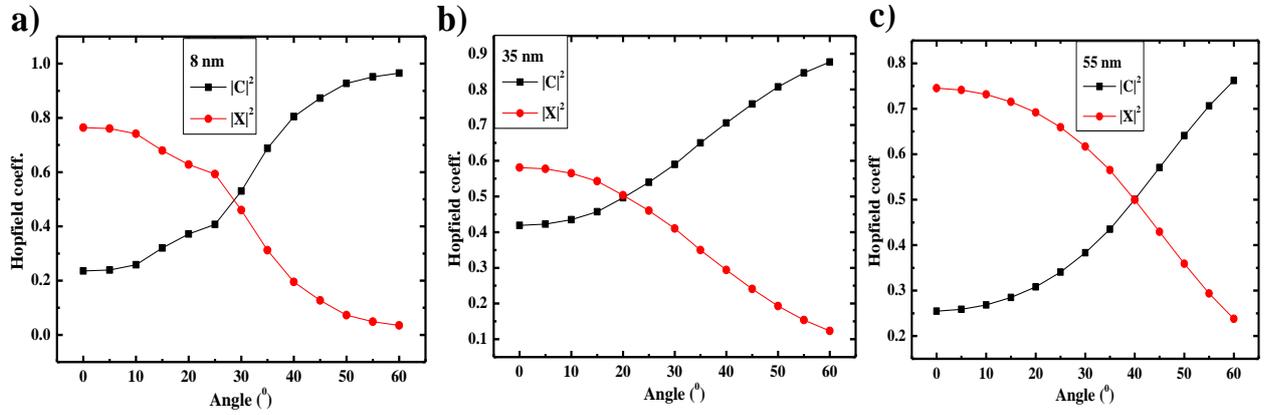

**Figure S11. Hopfield coefficient calculation for lower polariton branch:** (a) In λ/2-cavity for 8 nm thick PDI, (b) and (c) in λ-cavity for 35 nm and 55 nm thick PDI.

## Section-7

**Electrical measurements:** Electrical measurements were done inside the glove box under nitrogen atmosphere ($O_2$ and $H_2O$ level less than 10 ppm) using MB 150, Cascade Microtech probe station, and Keithley 2635 instrument controlled by Lab tracer software. Charge carrier mobility is calculated in the saturation regime using the following formula at fixed drain voltage ($V_D = 40$ V):

$$\mu = \left(\frac{d\sqrt{I_D}}{dV_G}\right)^2 \frac{2L}{WC_i}$$

where L (Length of the gold electrodes) = 20 µm, W (width of the electrodes) = 2 mm, $C_i$ (capacitance of gate dielectric) = 3.84 x $10^{-8}$ F cm$^{-2}$ (for 90 nm $SiO_2$) and 1.5 x $10^{-8}$ F cm$^{-2}$ (for 230 nm $SiO_2$).

Some of the experimental plots are shown below; the first column shows output characteristics plots ($I_D$-$V_D$), and the second column shows transfer characteristics plots ($I_D$-$V_D$) for various samples.

## I. For λ/2-Cavity:

| | 1) 0.05% PDI | |
|---|---|---|
| Sr. no. | $I_D$-$V_D$ | $I_D$-$V_G$ |
| 1. | | |
| 2. | | |
| 3. | | |

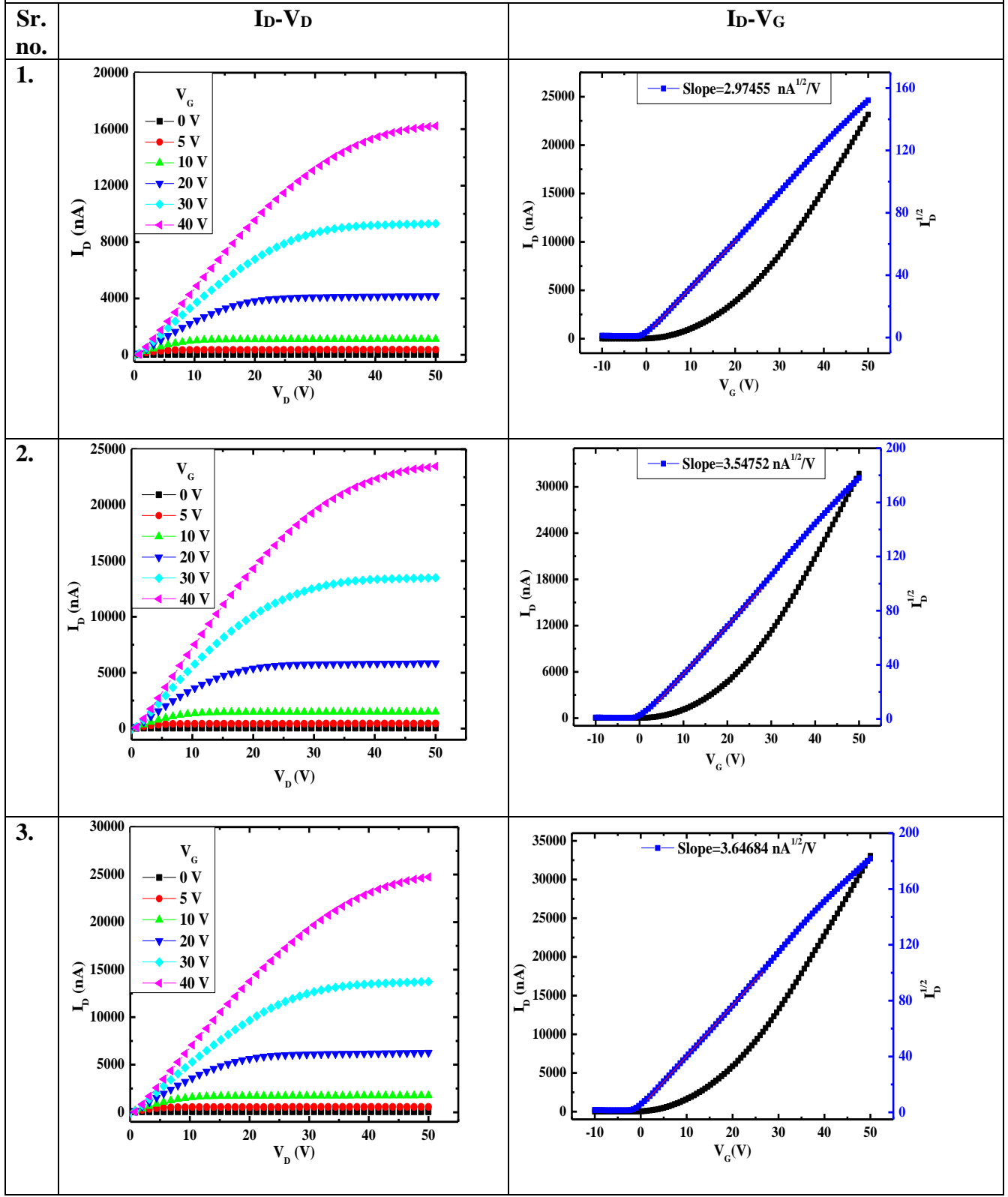

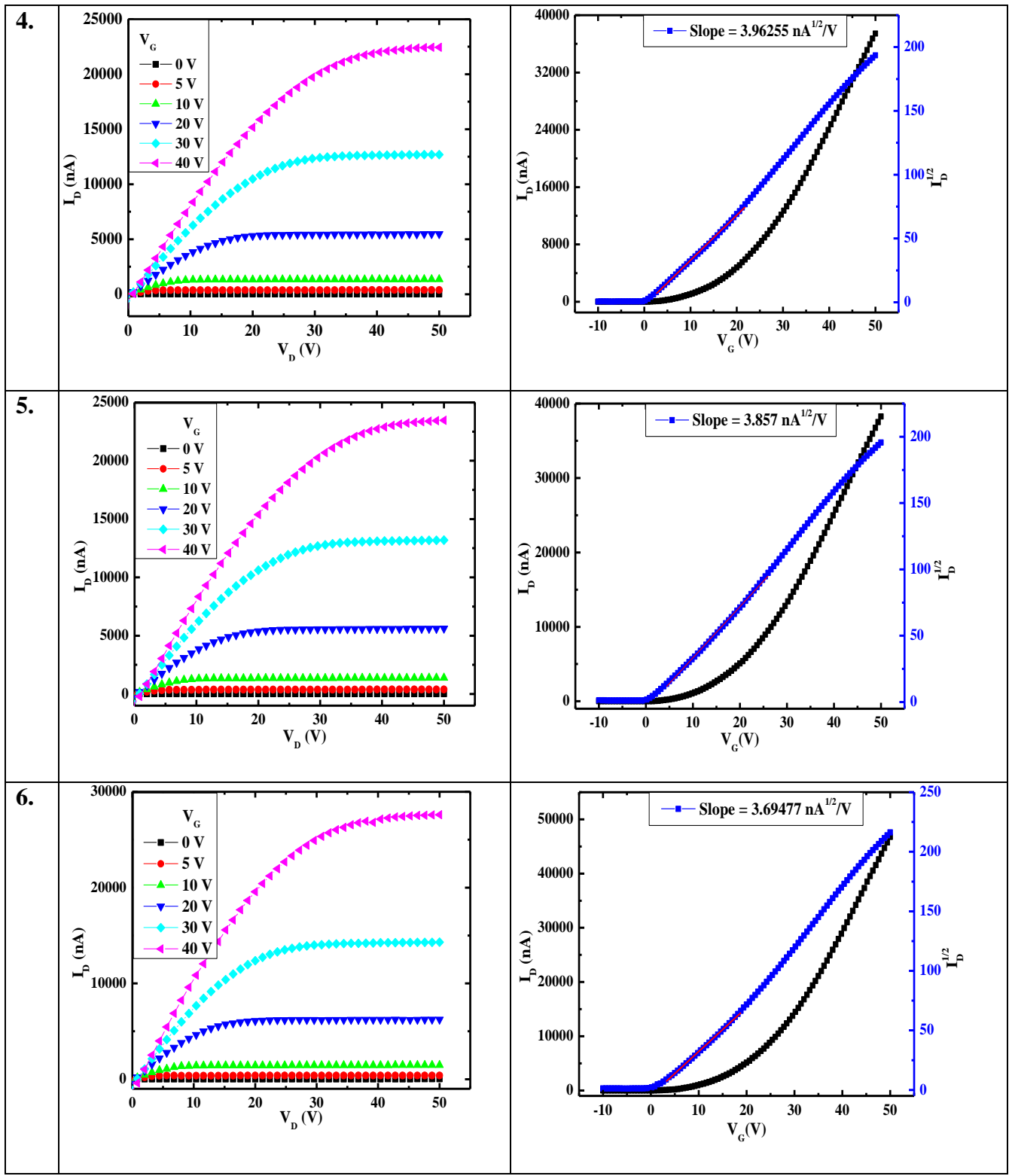

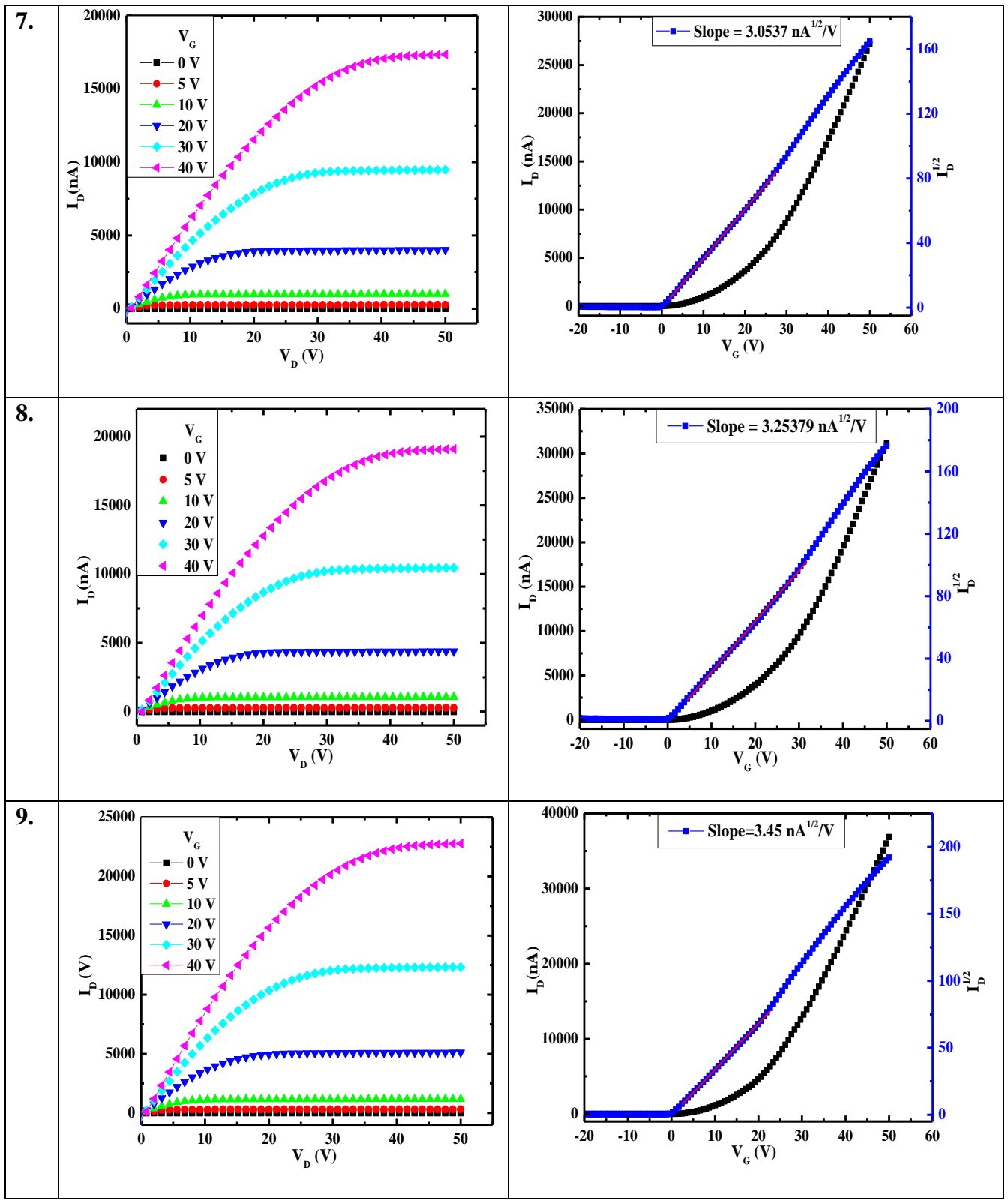

| 10. | 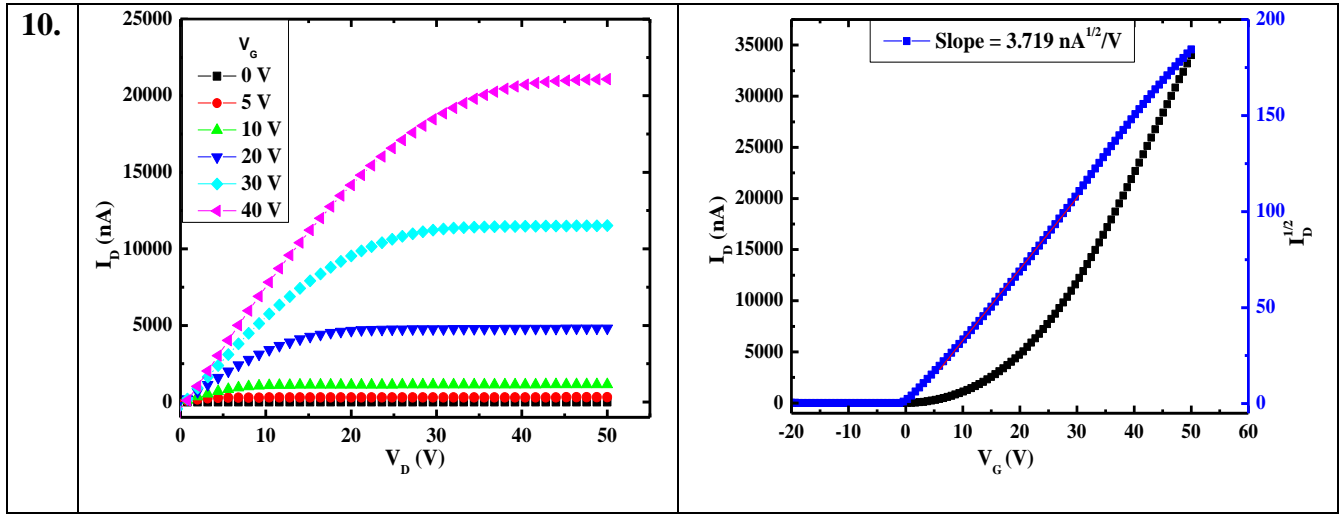 |

| | **2)   0.1% PDI** |
|---|---|
| Sr. no. | $I_D$-$V_D$ ... $I_D$-$V_G$ |
| 1. 2. | 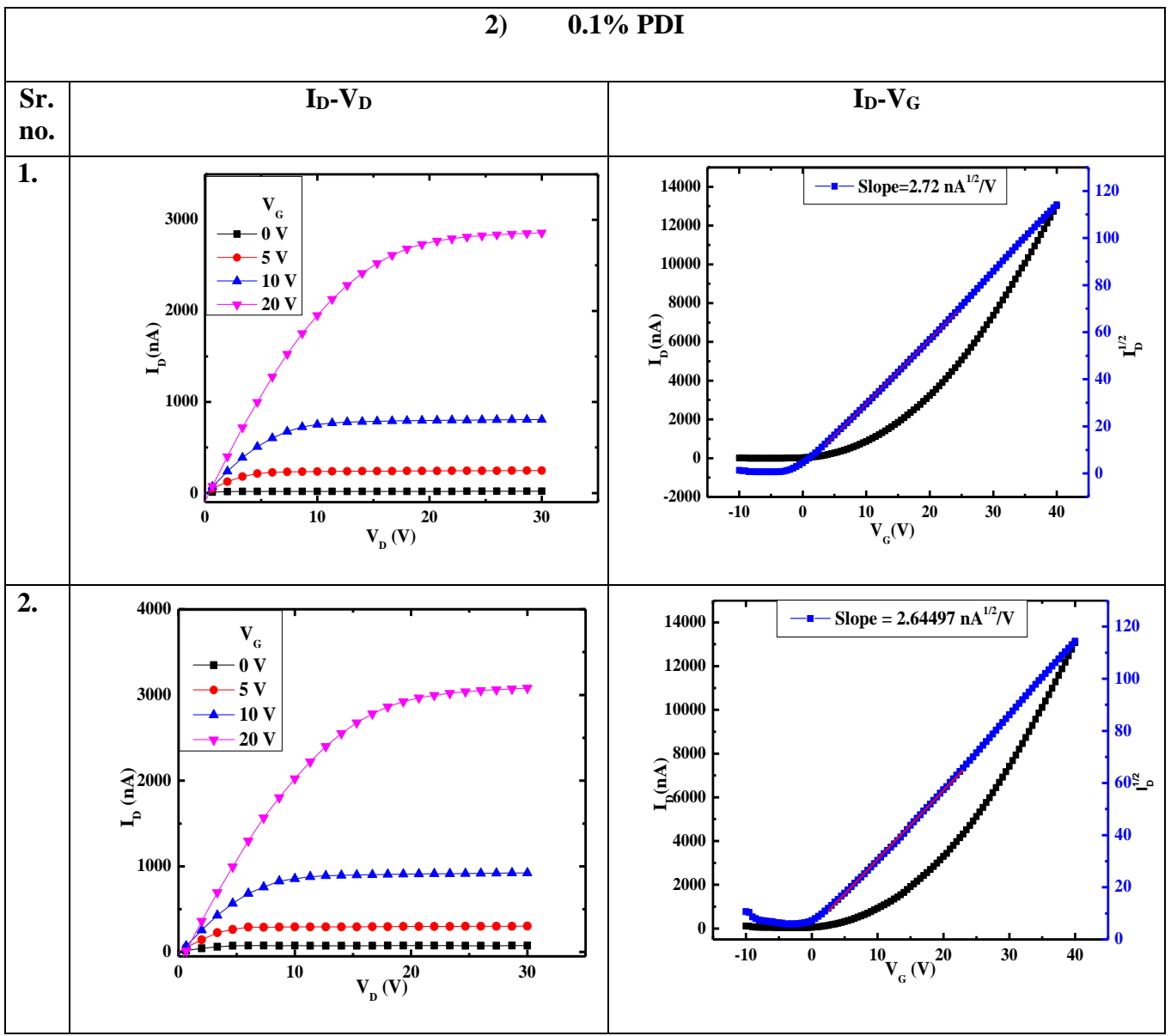 |

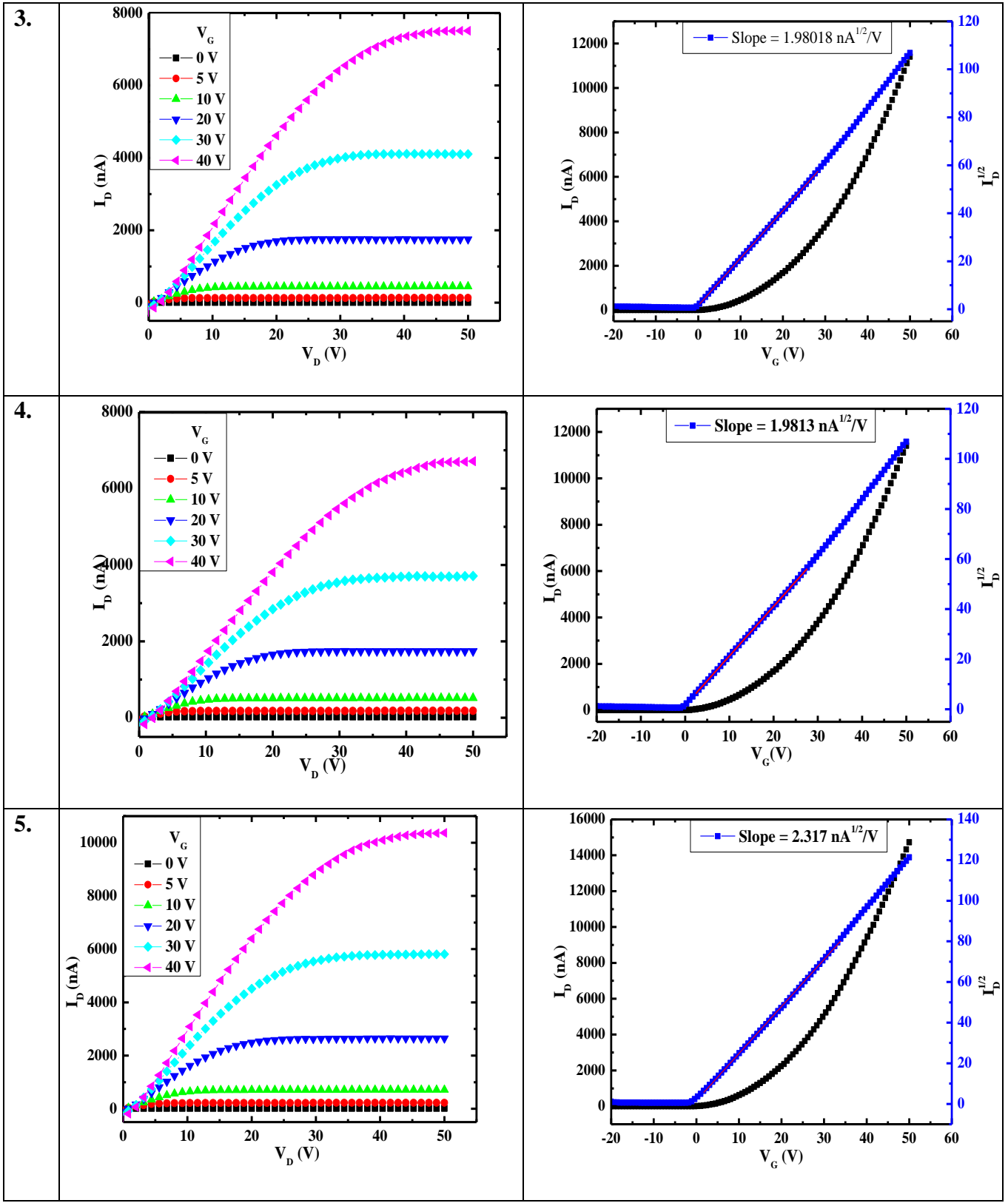

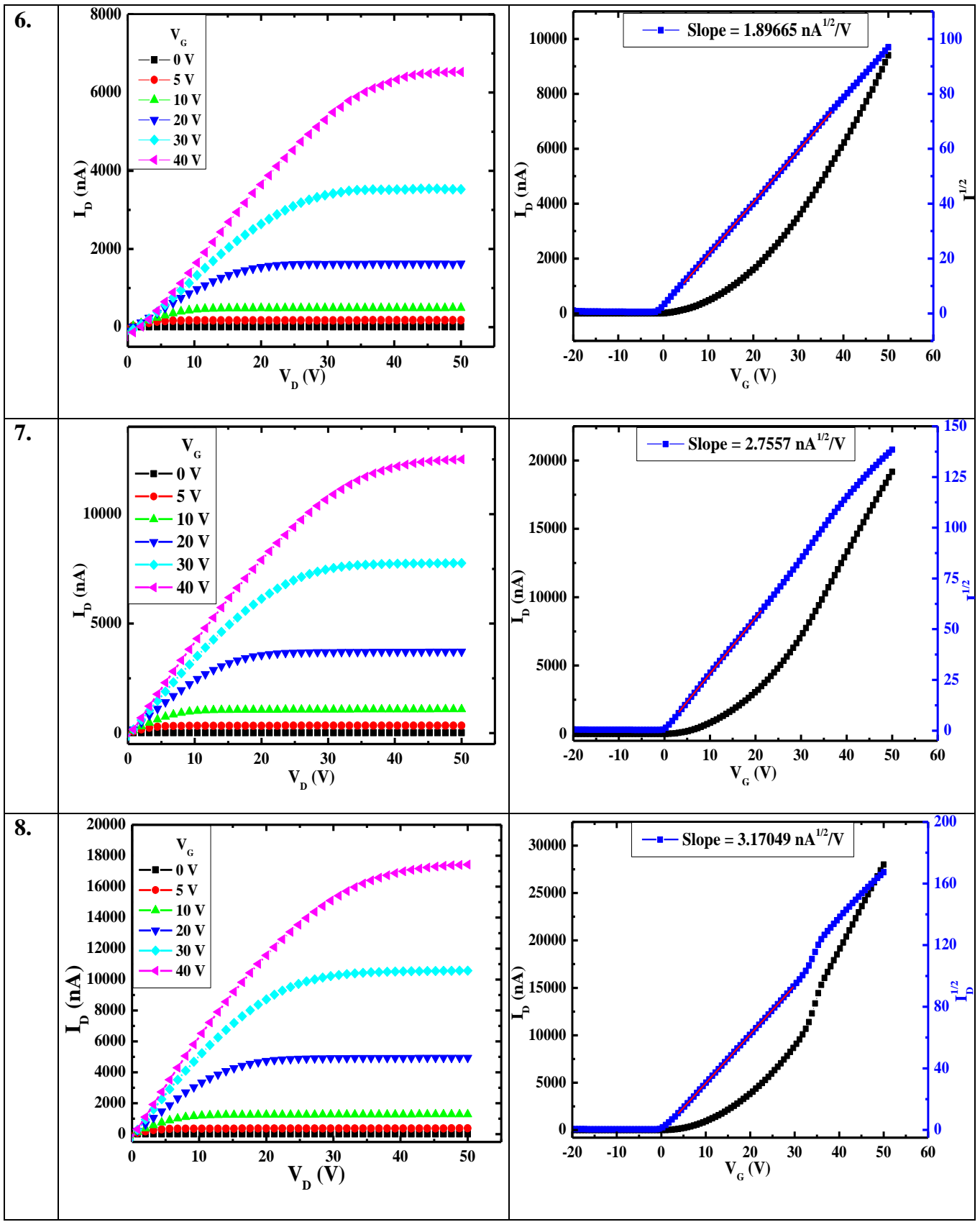

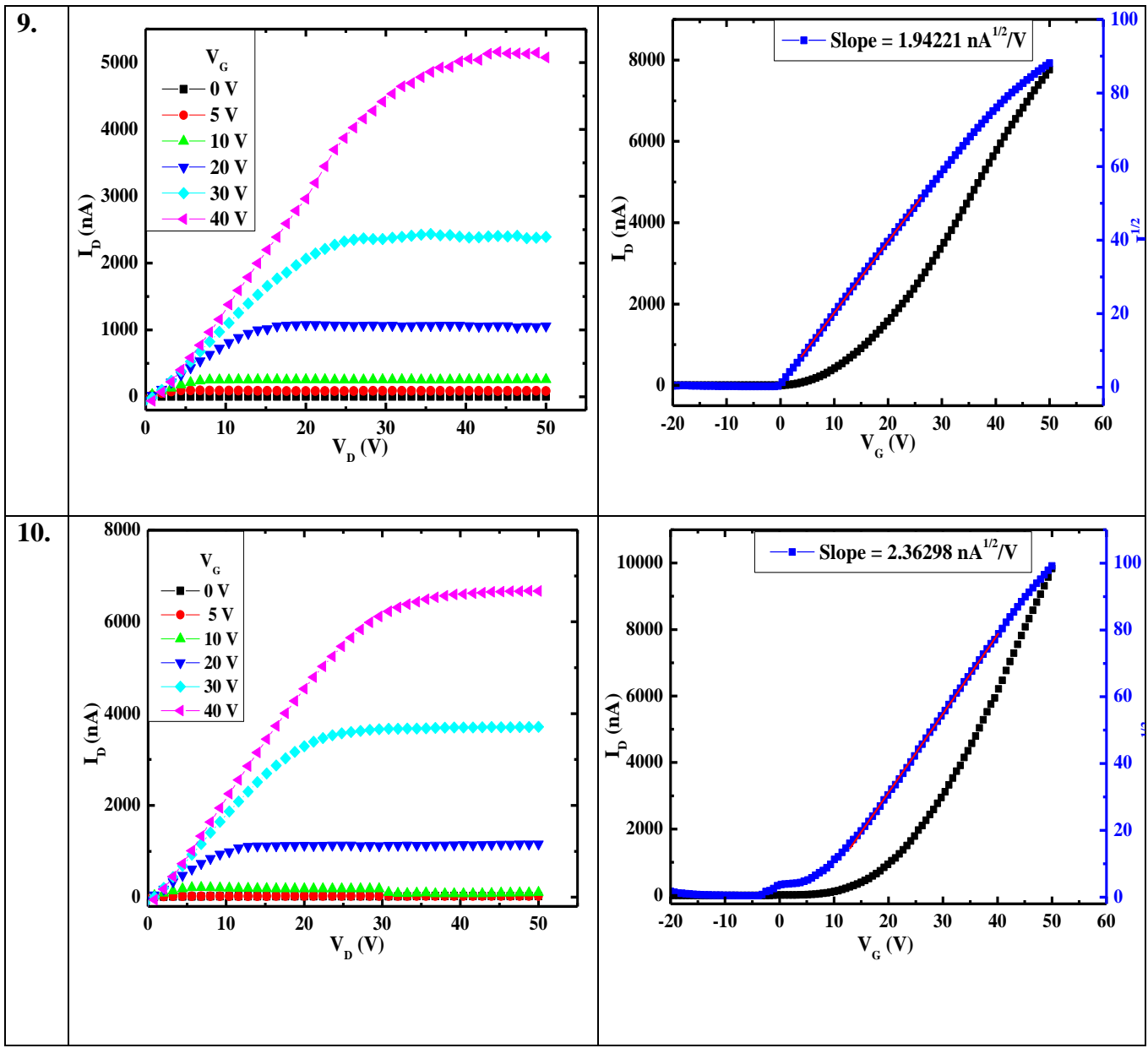

| 3) 0.2% PDI | |
|---|---|
| Sr. no. | $I_D$-$V_D$ | $I_D$-$V_G$ |

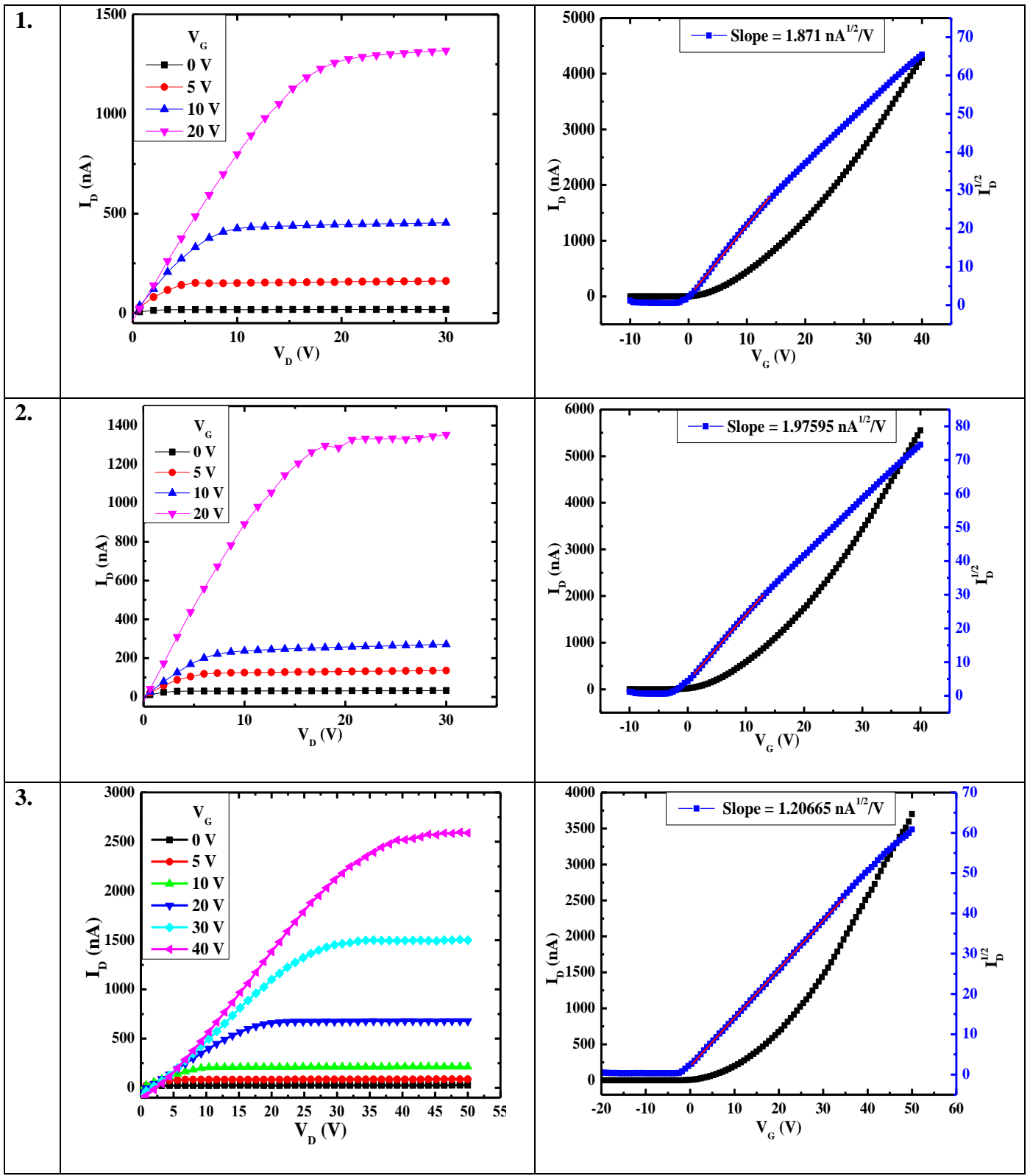

| 4. | 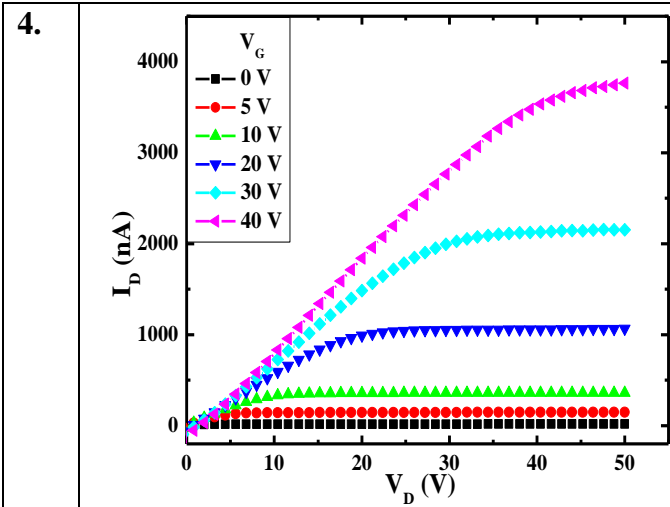 | 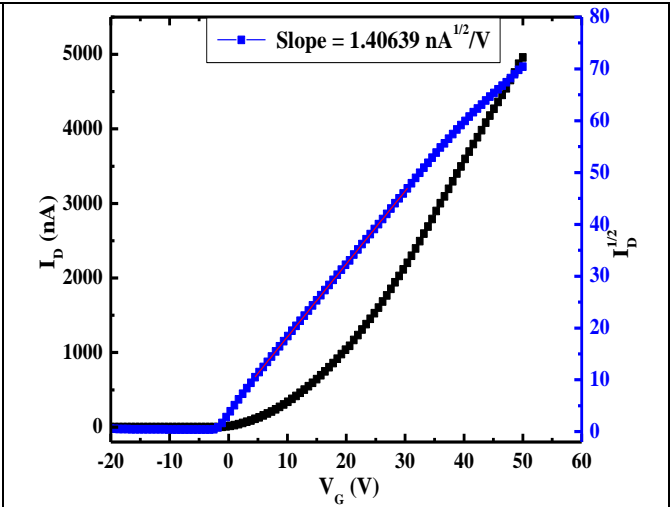 |
|---|---|---|
| 5. | 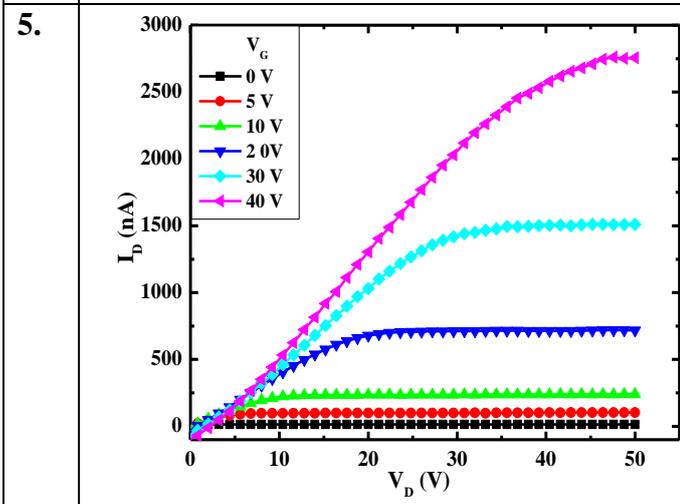 | 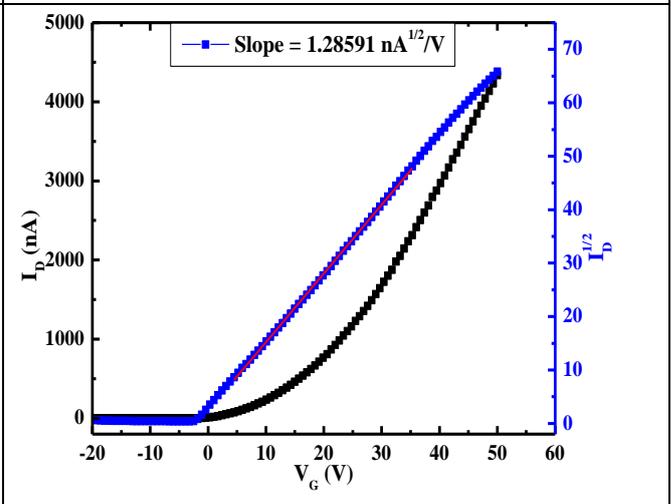 |
| 6. | 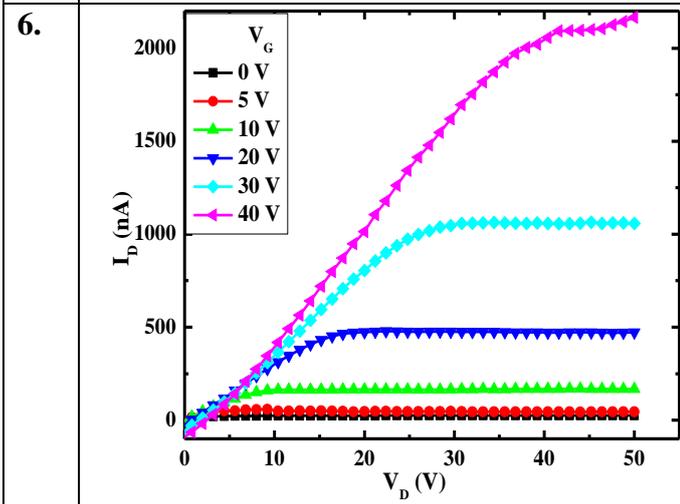 | 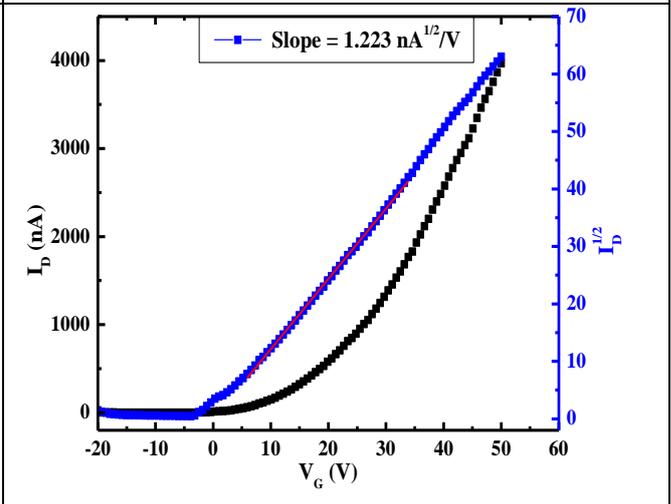 |

| 7. | 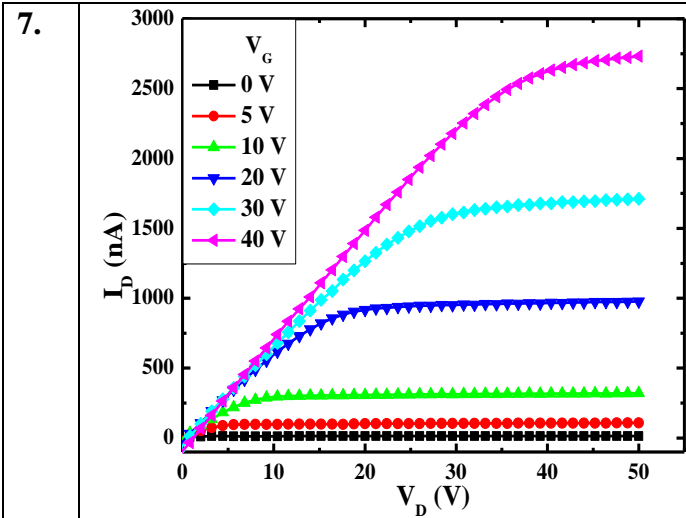 | 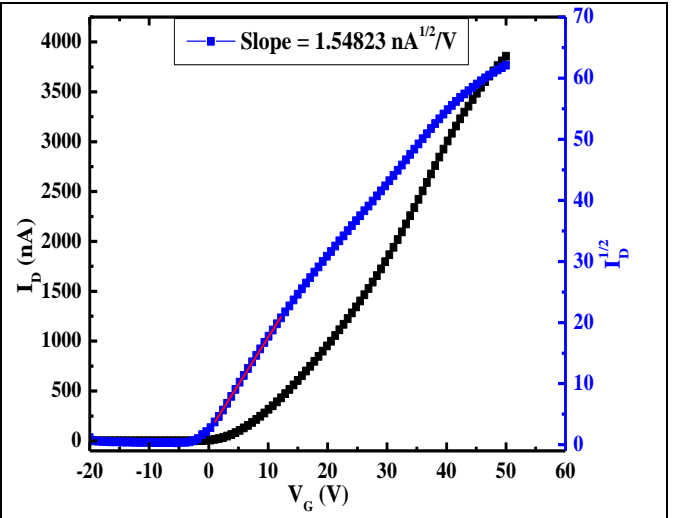 |
|---|---|---|
| 8. | 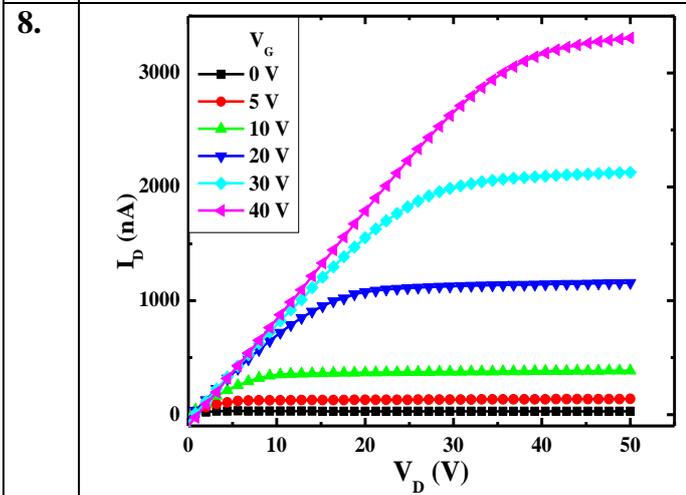 | 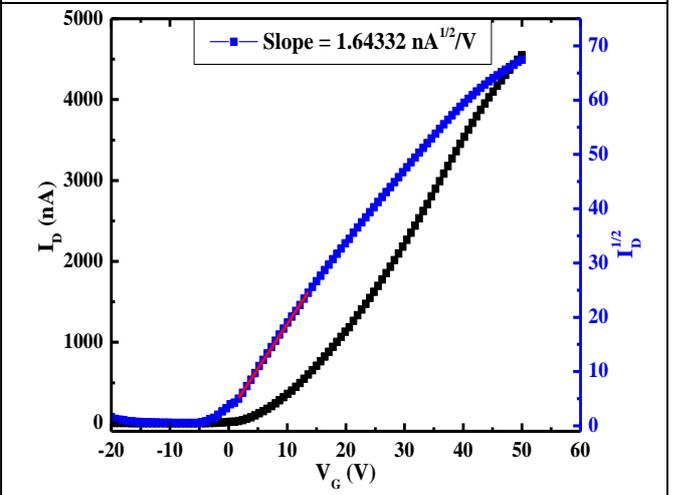 |
| 9. | 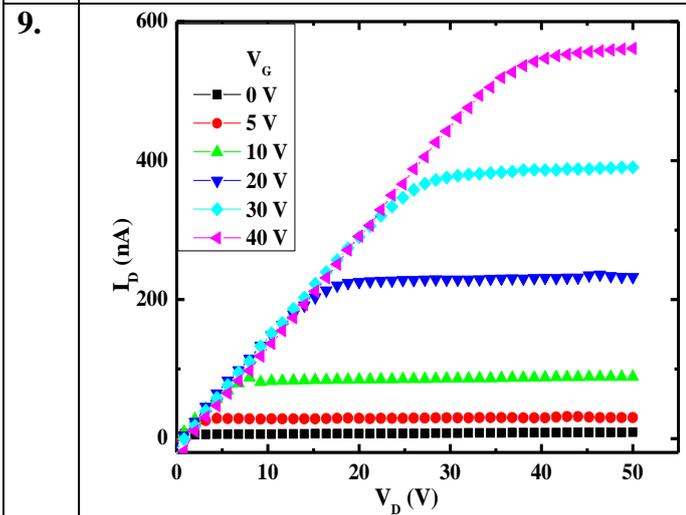 | 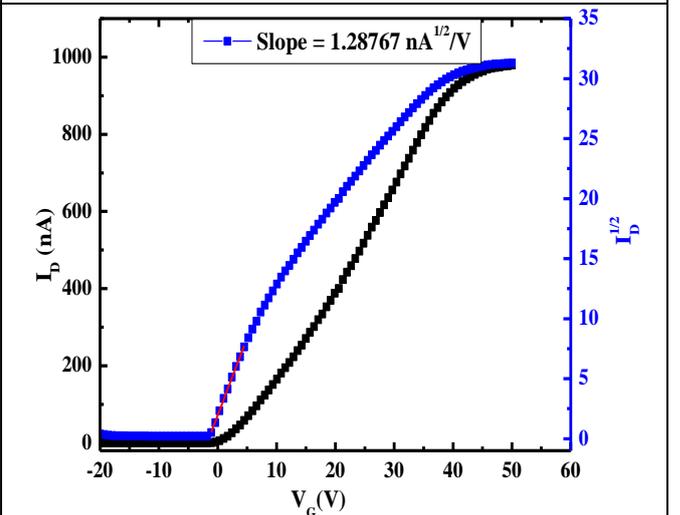 |

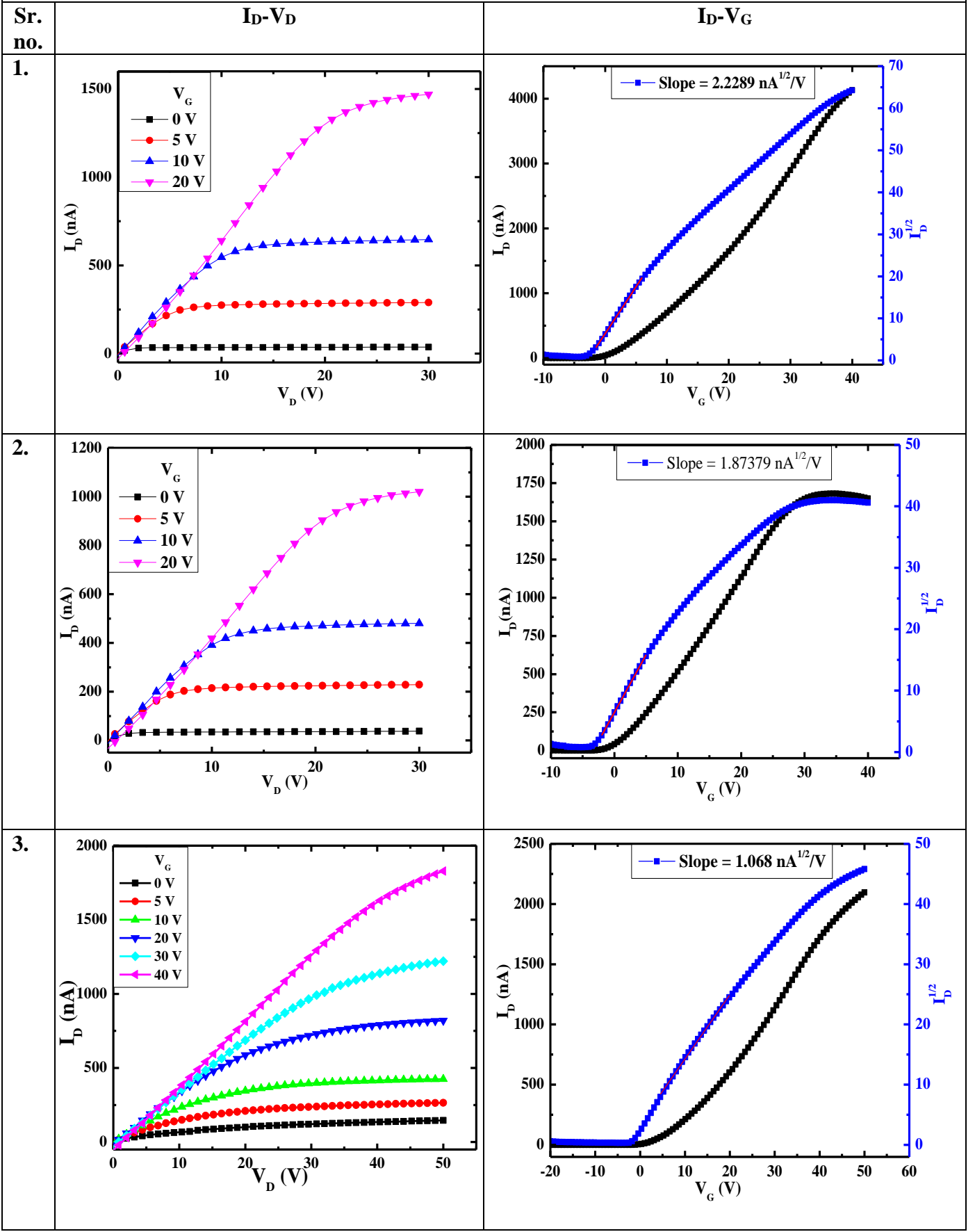

4.

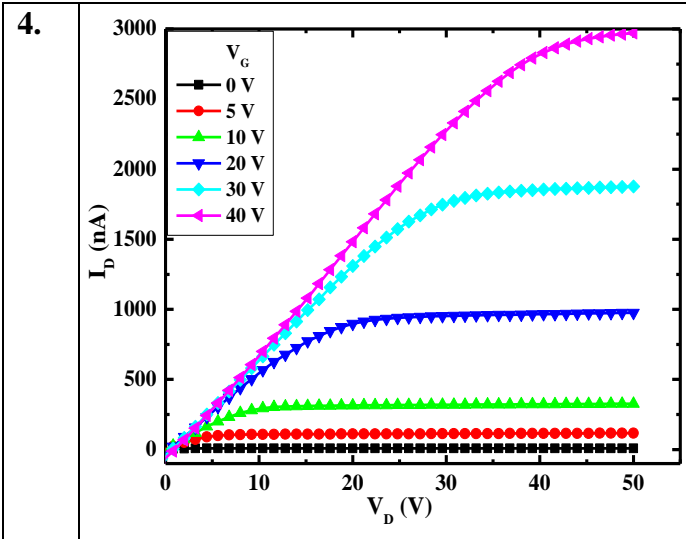 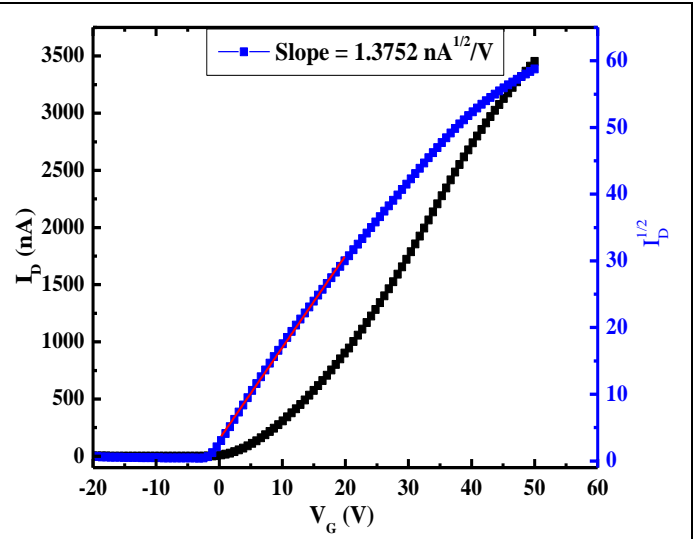

5.

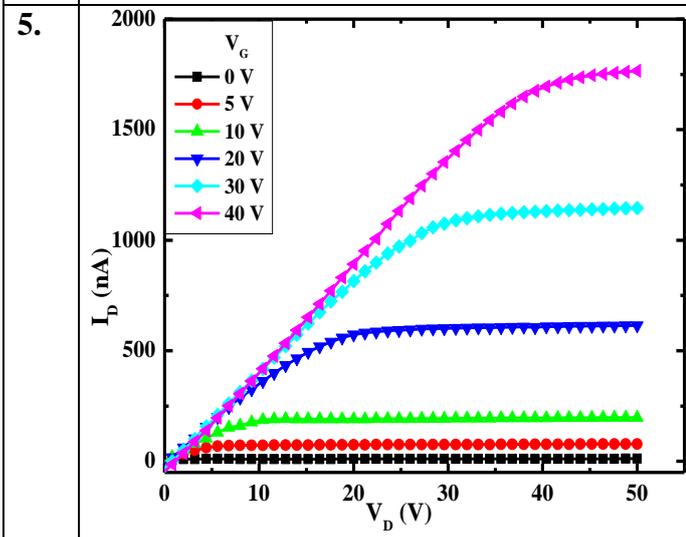 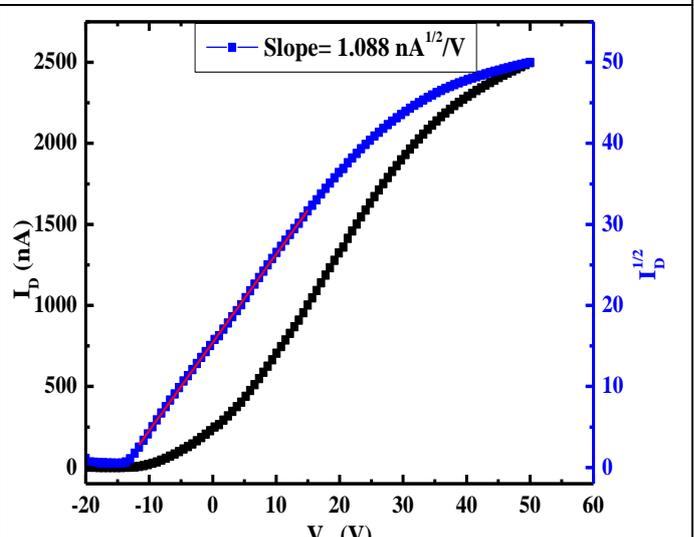

6.

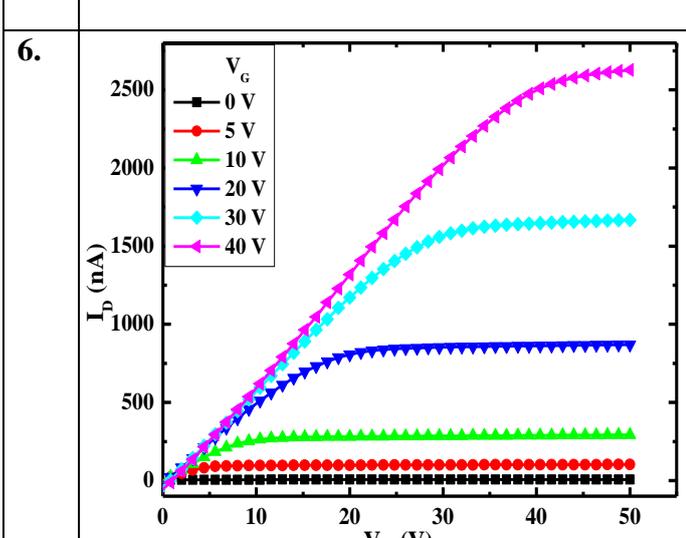 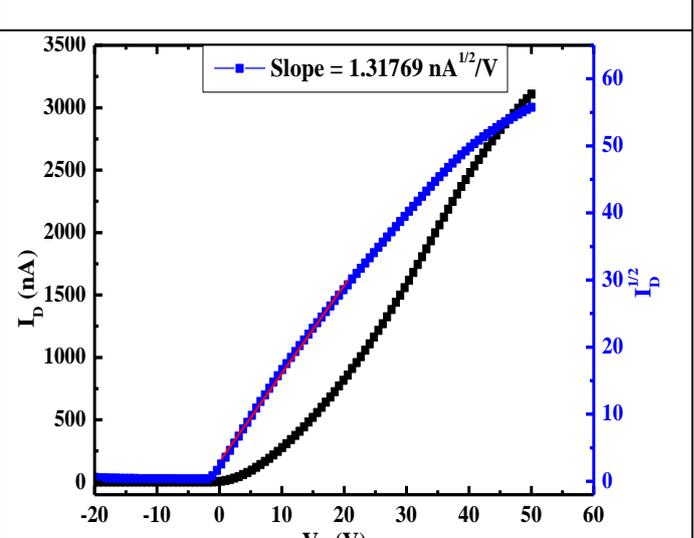

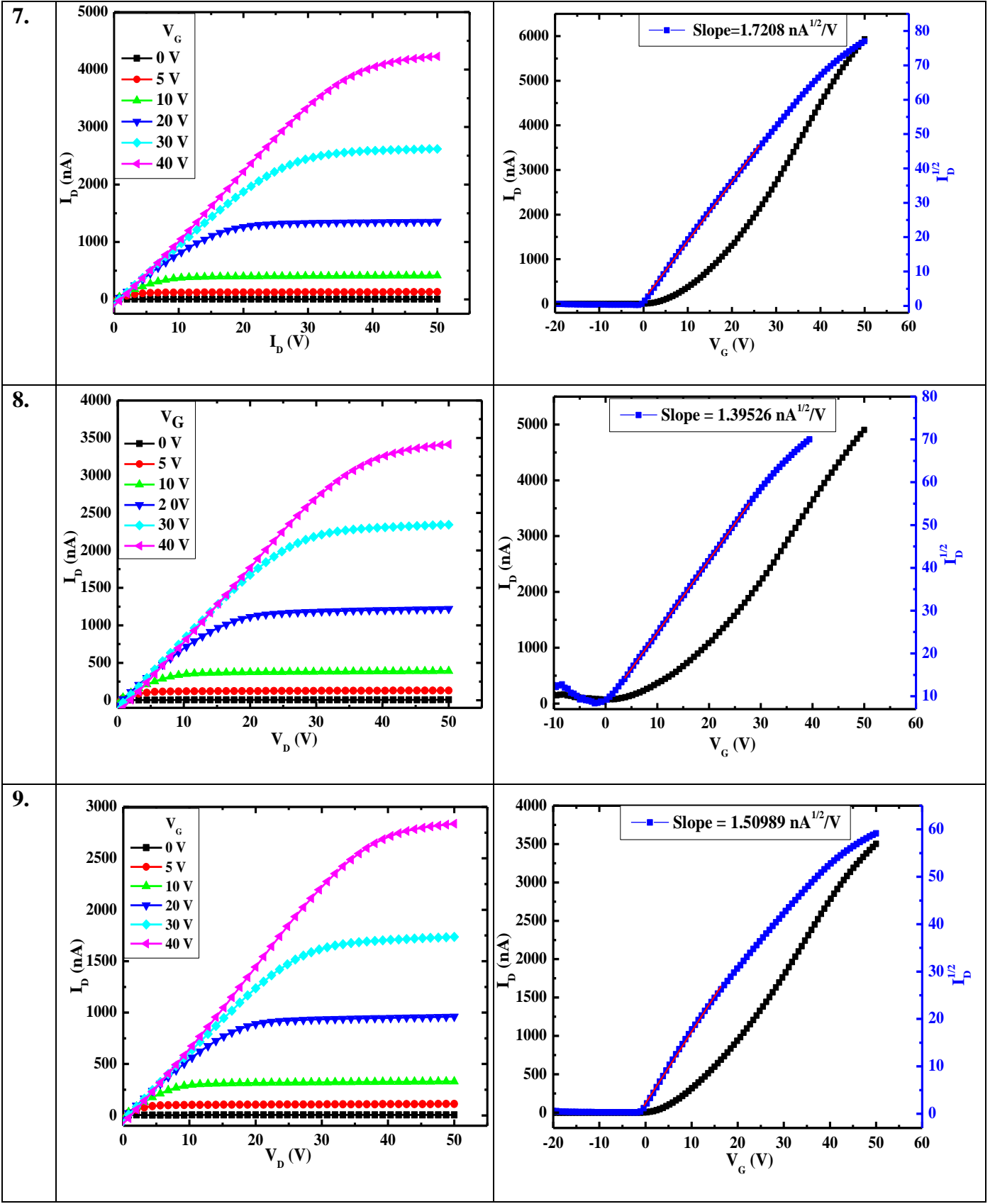

| 10. | 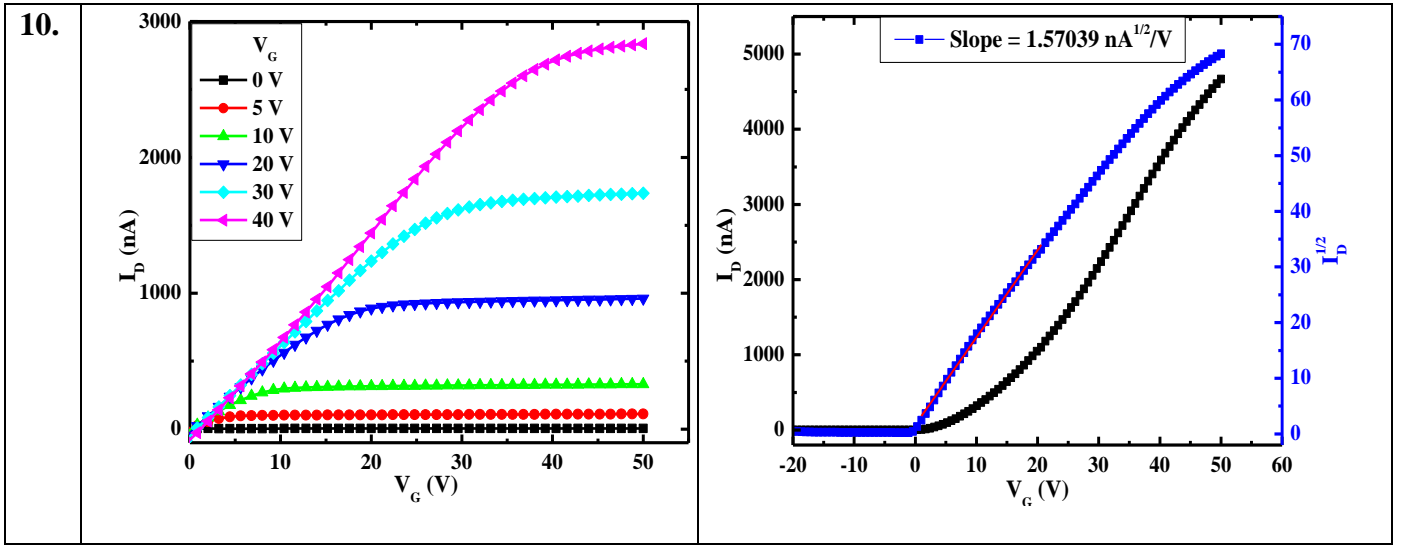 |

| 5) 0.5% PDI | |
|---|---|
| Sr. no. | $I_D$-$V_D$ | $I_D$-$V_G$ |

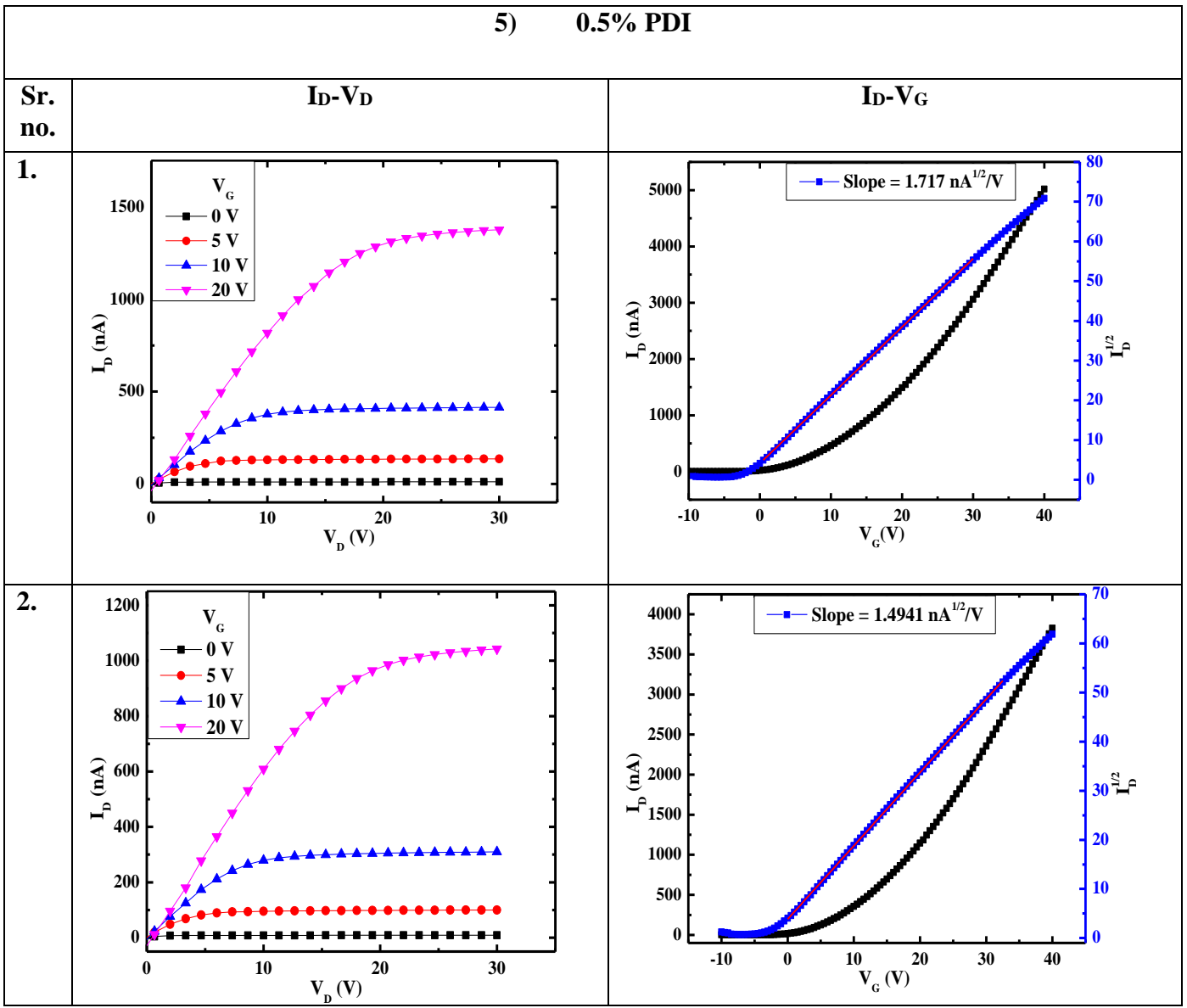

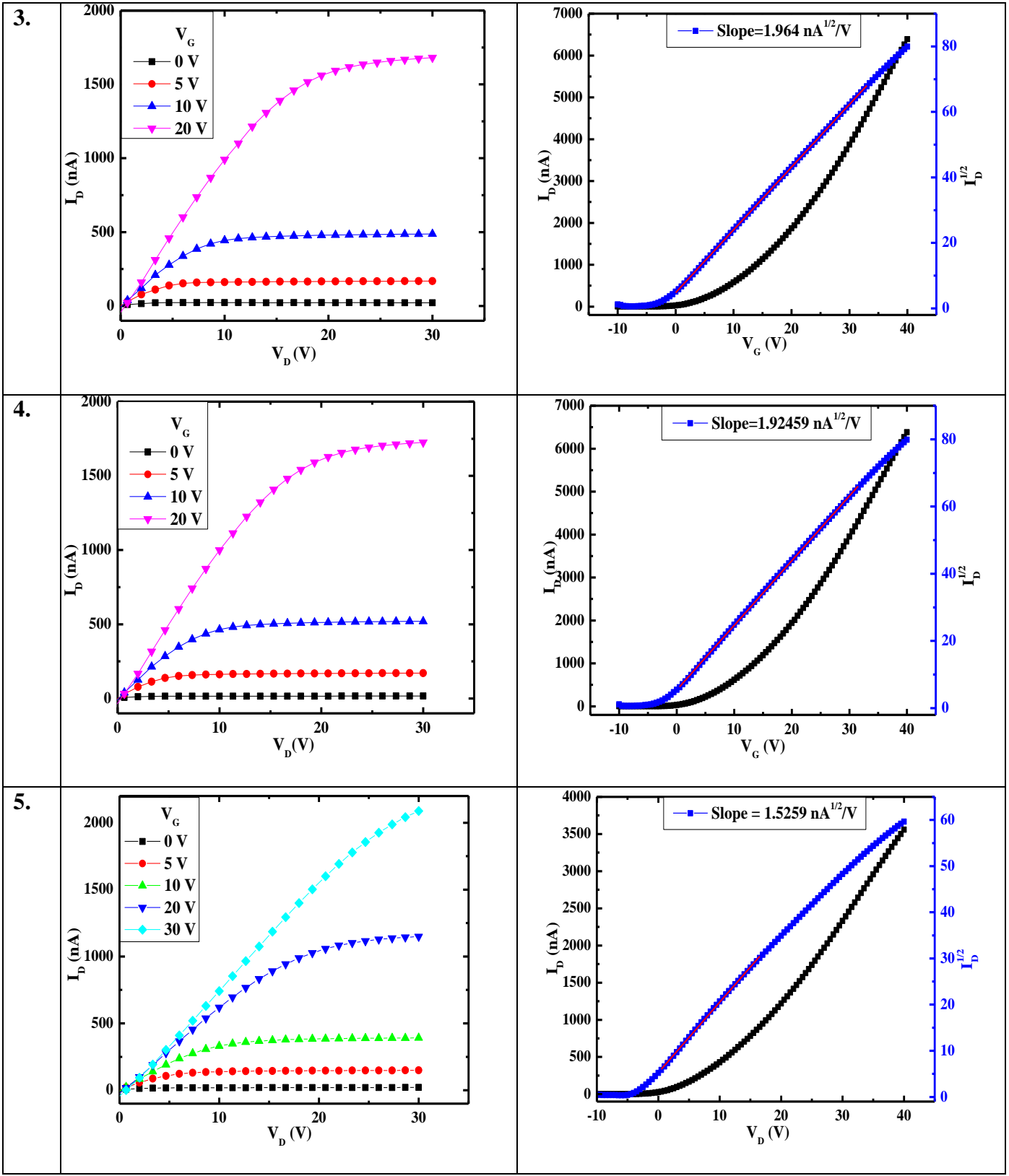

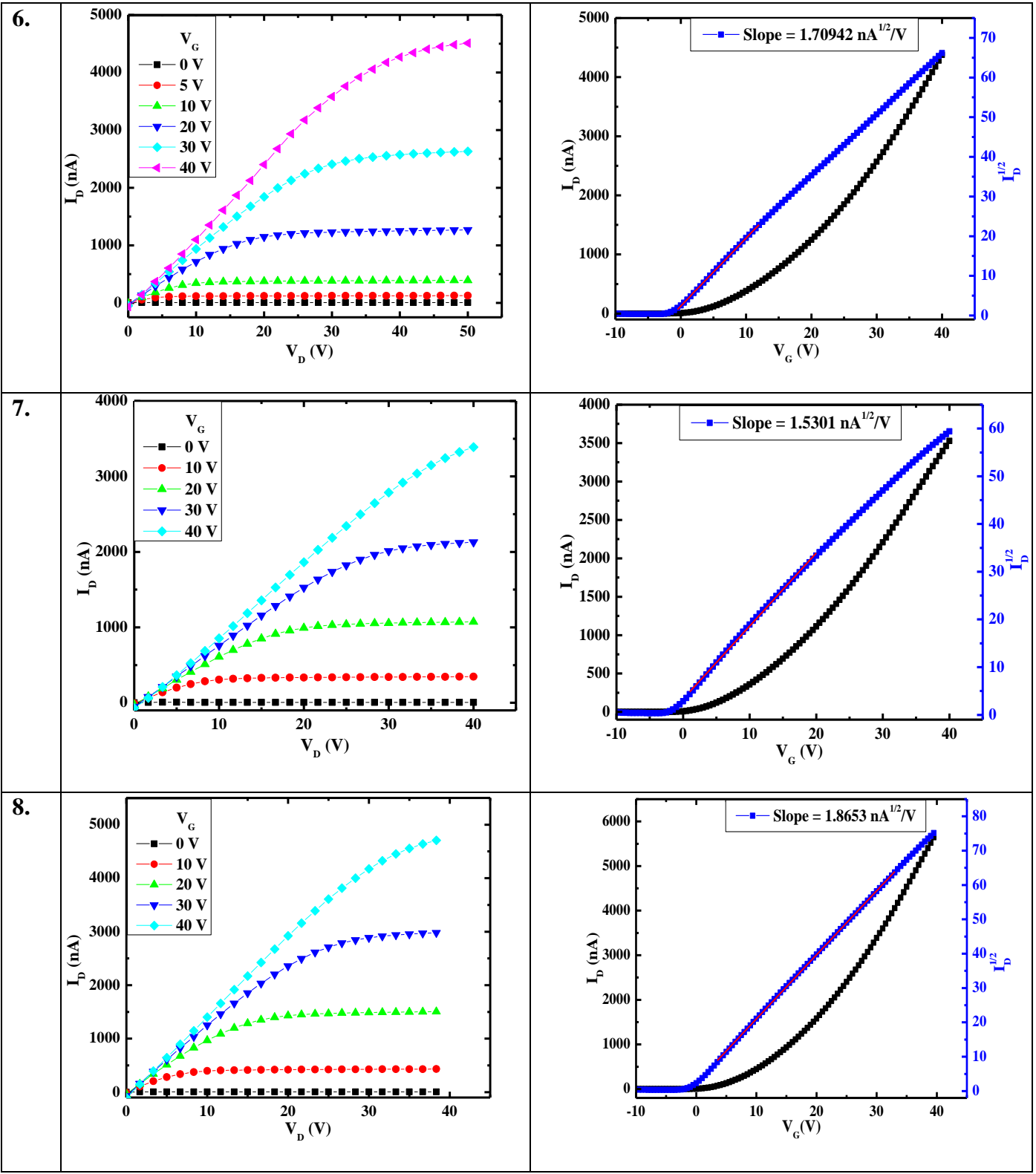

| 9. | 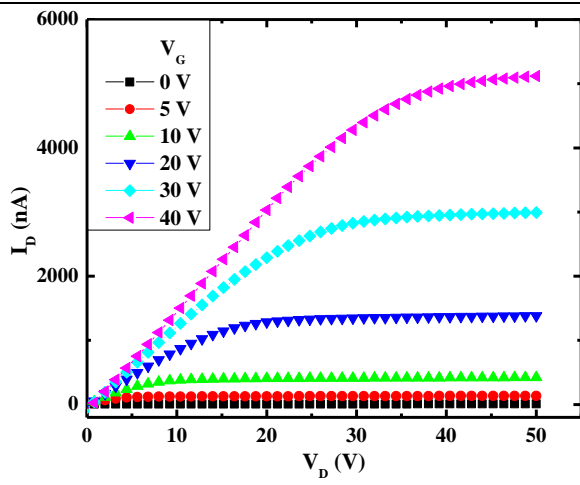 | 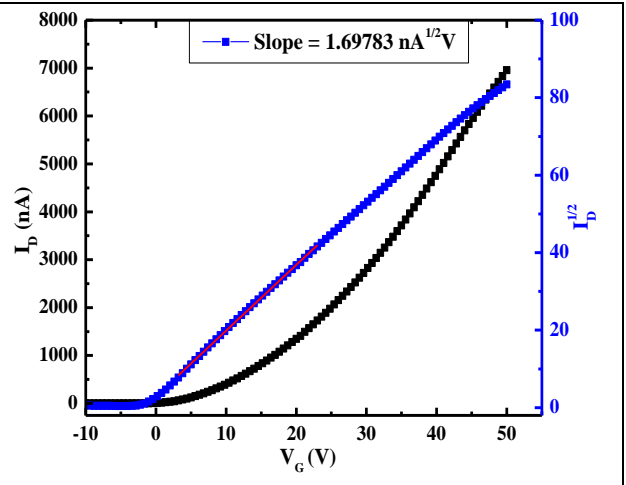 |
|---|---|---|
| 10. | 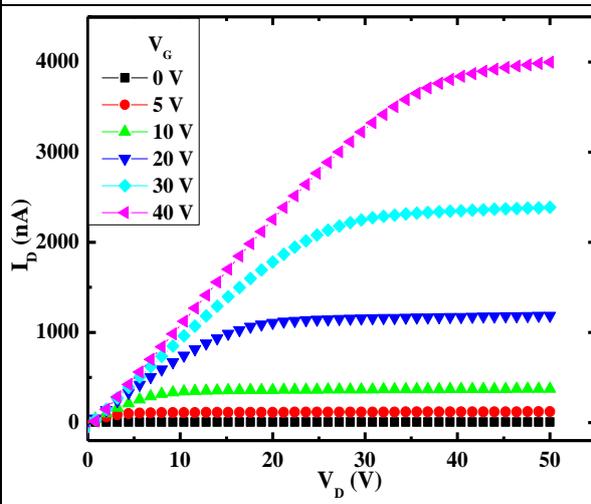 | 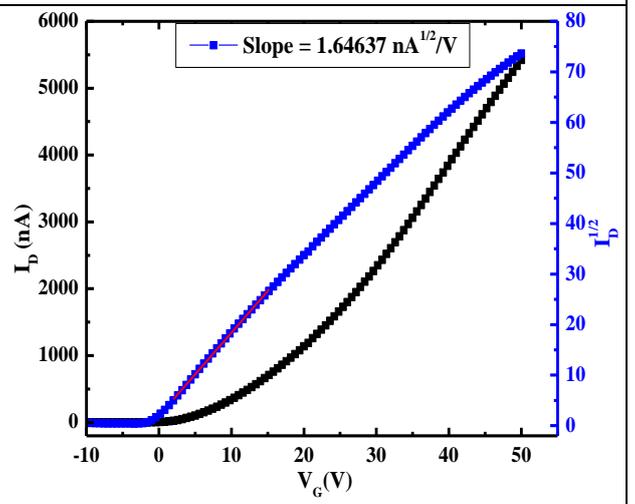 |

## II. For λ-Cavity

| | 1) 0.05% PDI | |
|---|---|---|
| Sr. no. | $I_D$-$V_D$ | $I_D$-$V_G$ |

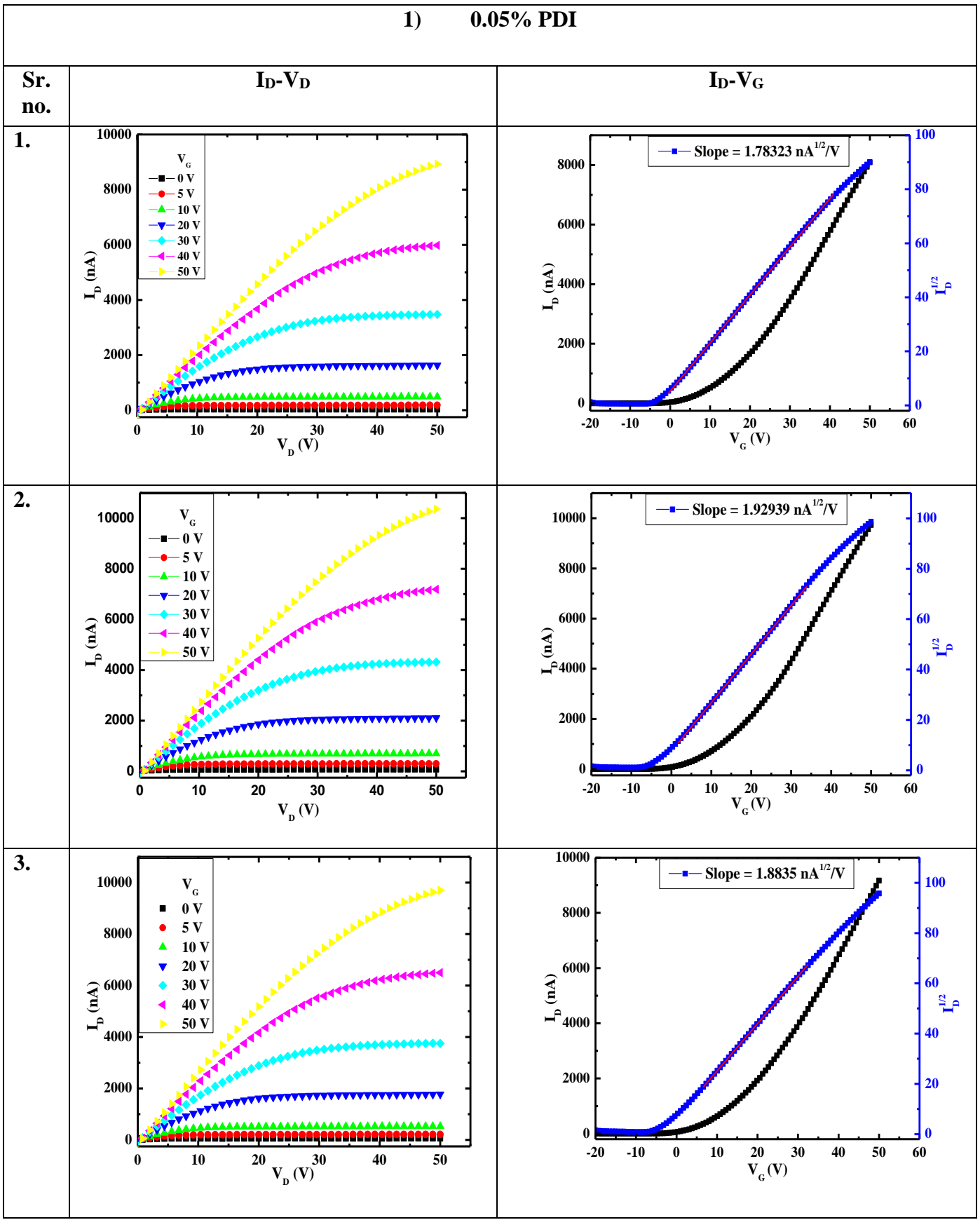

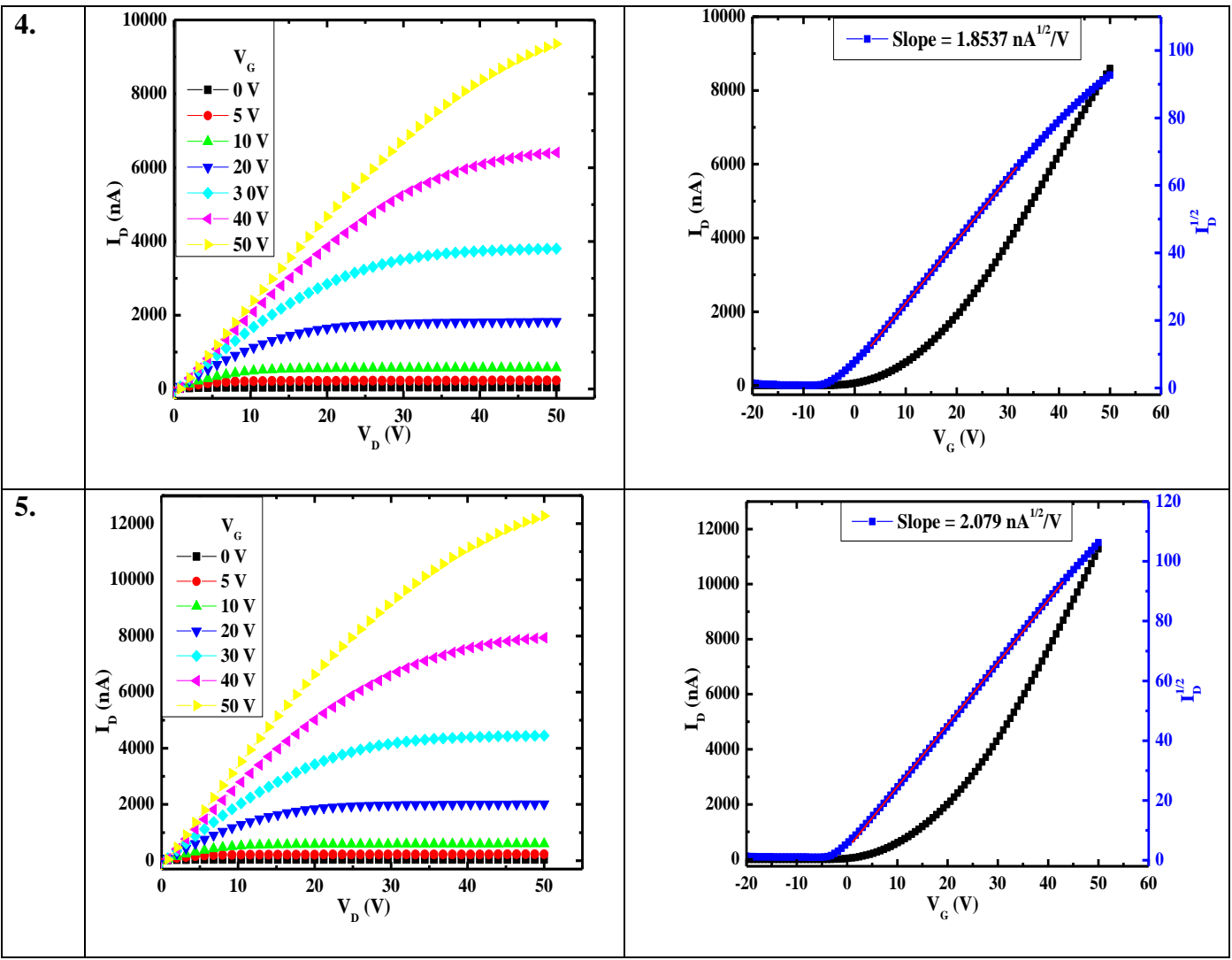

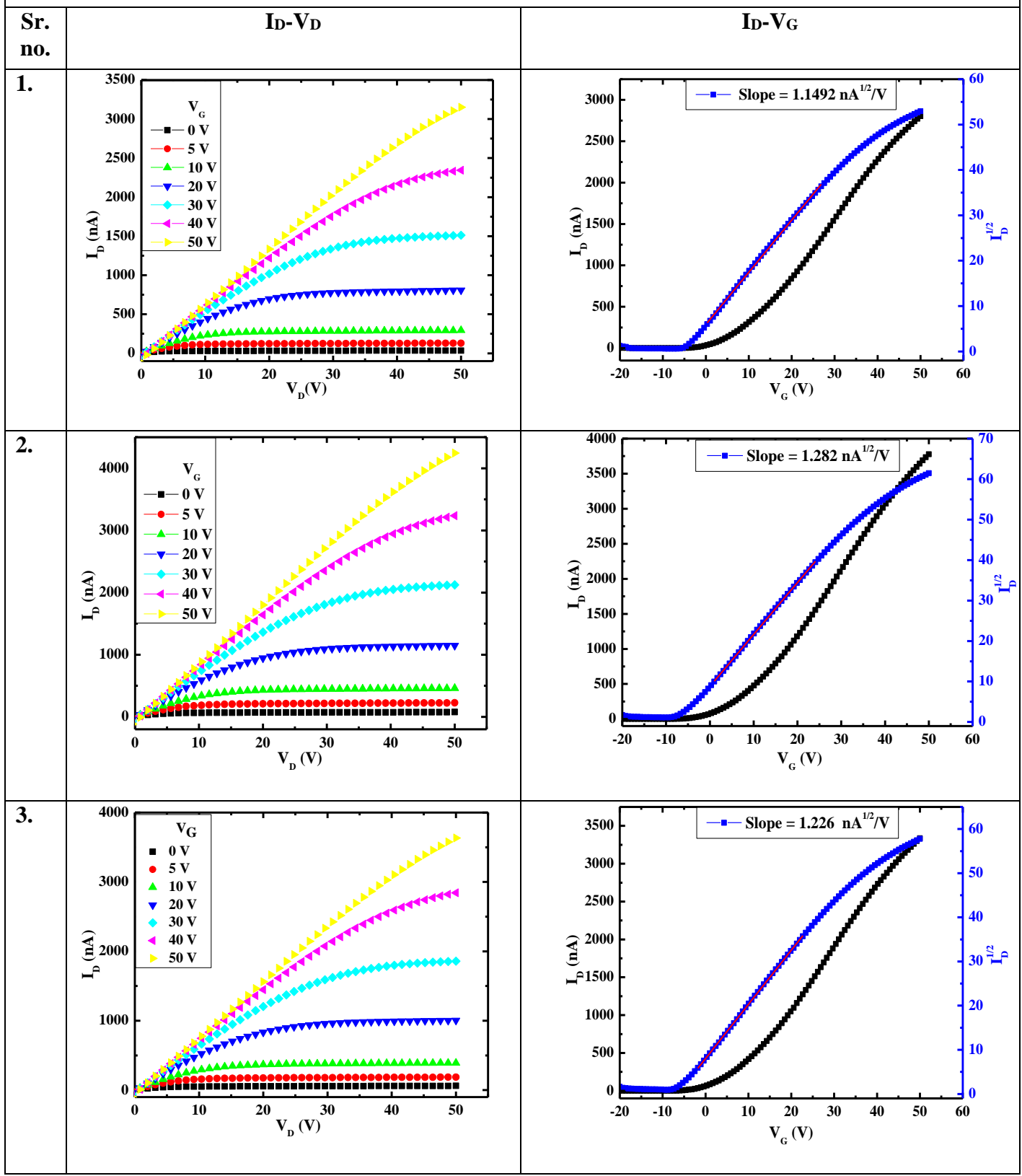

4. 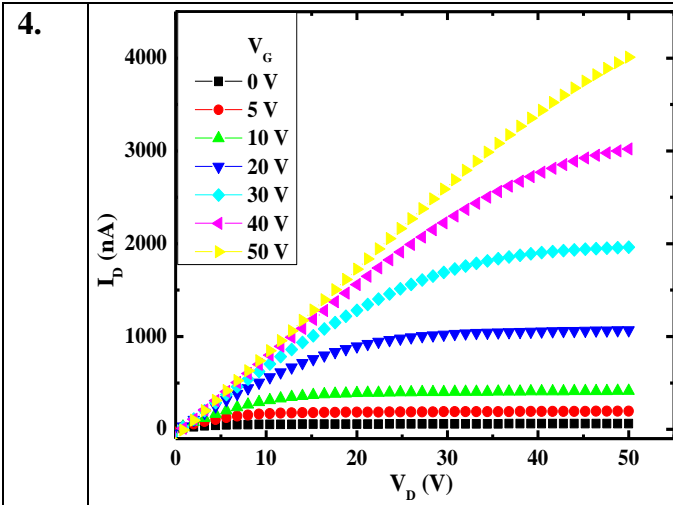 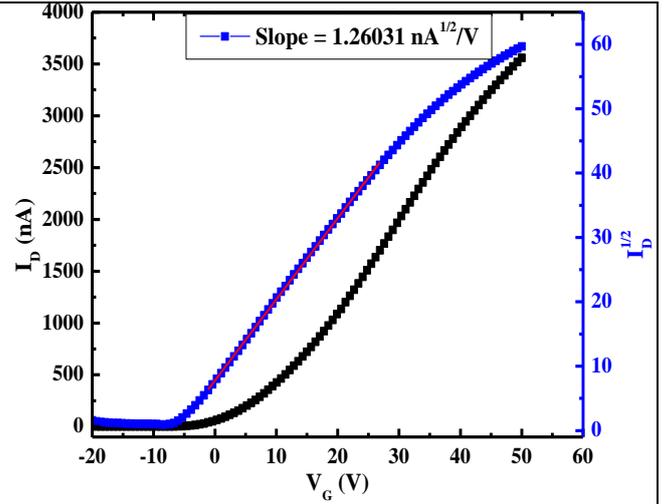

5. 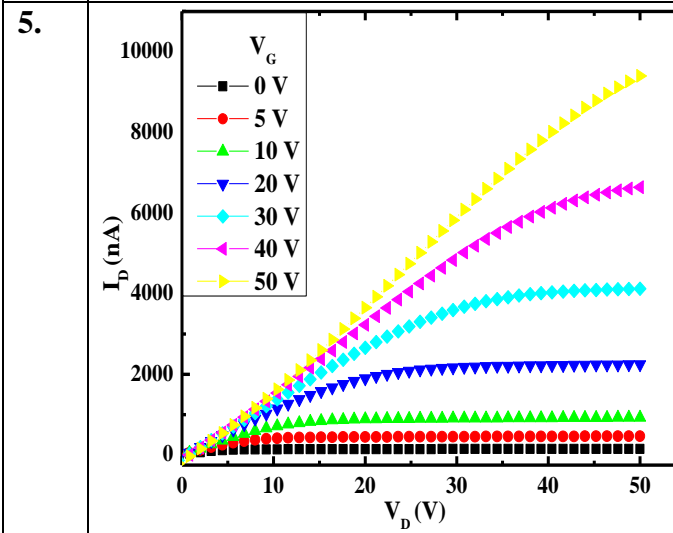 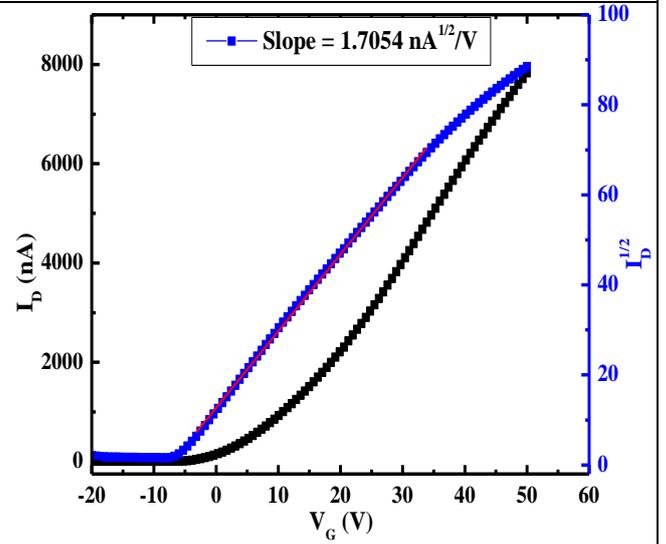

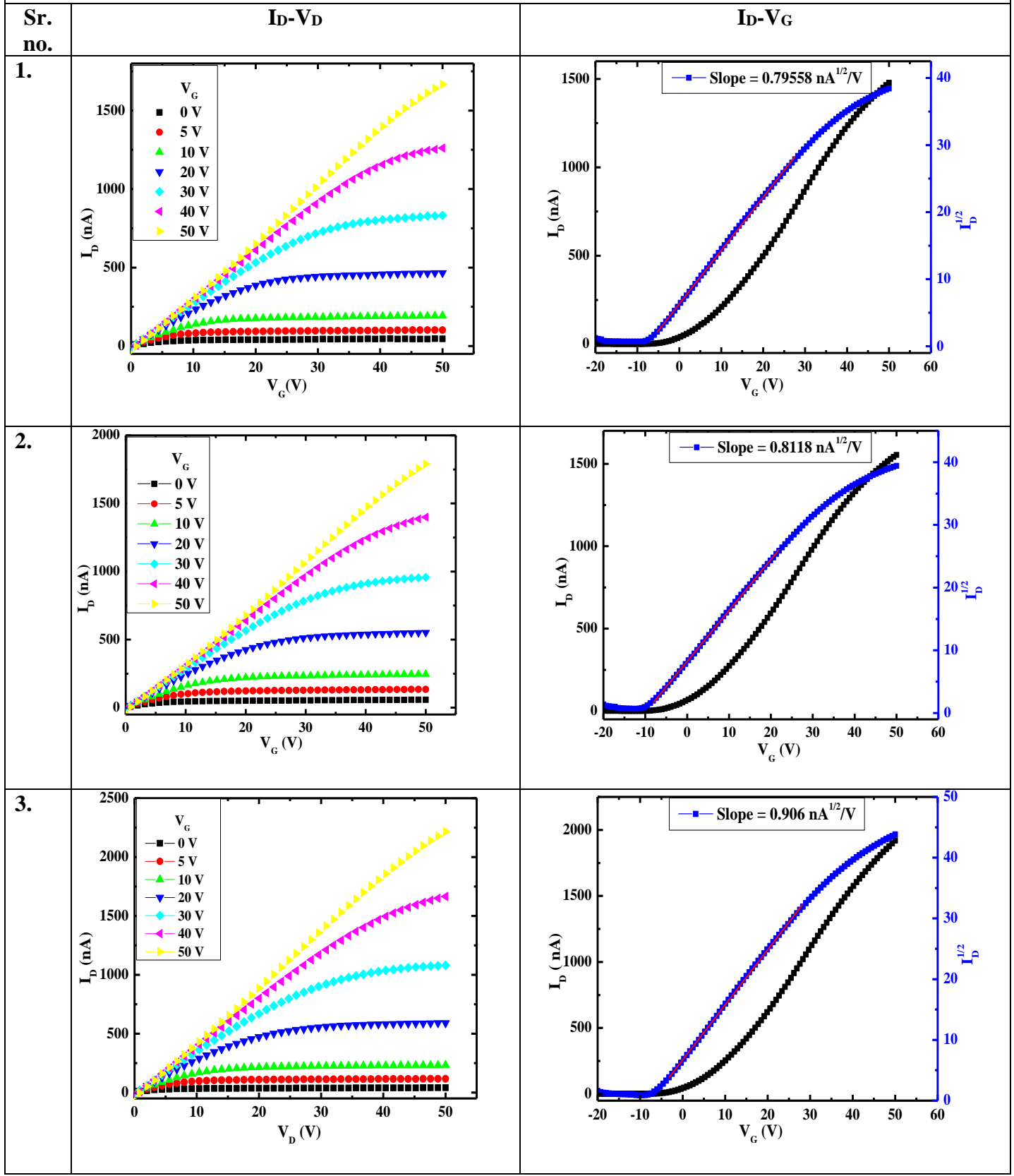

| 4. | 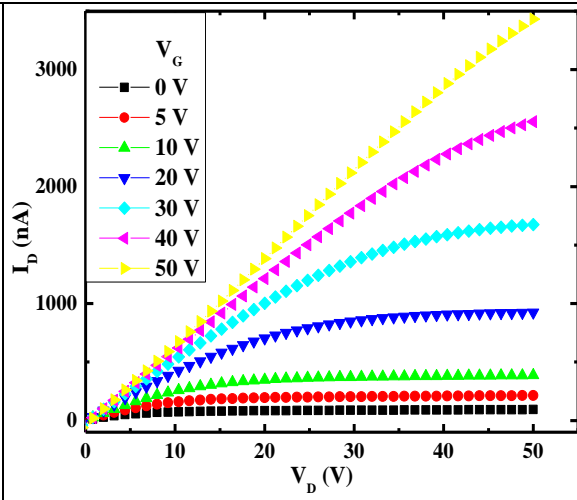 | 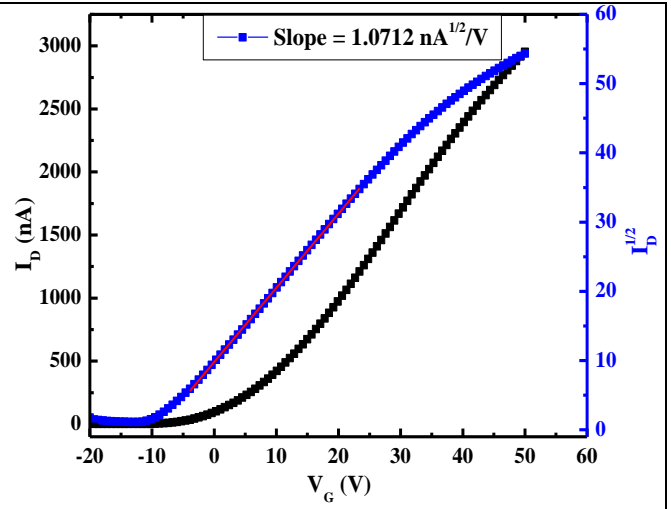 |
|---|---|---|
| 5. | 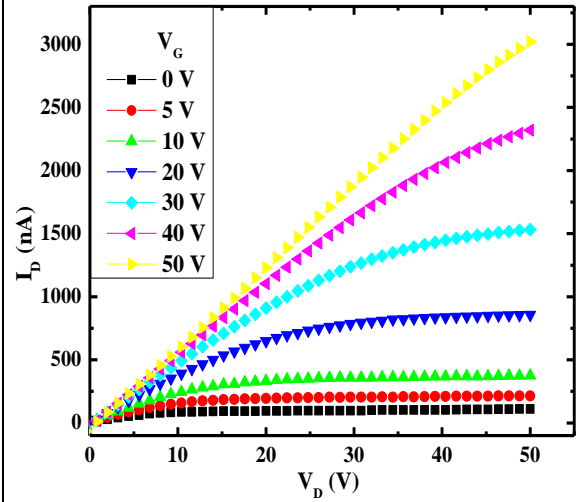 | 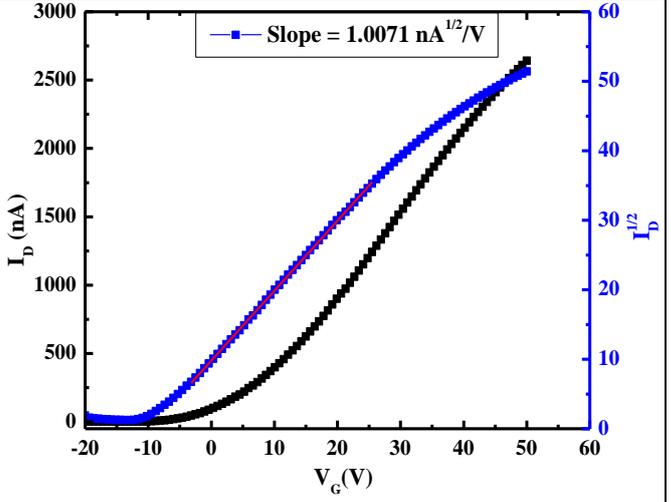 |

| | **4) 0.3% PDI** | |
|---|---|---|
| Sr. no. | $I_D$-$V_D$ | $I_D$-$V_G$ |

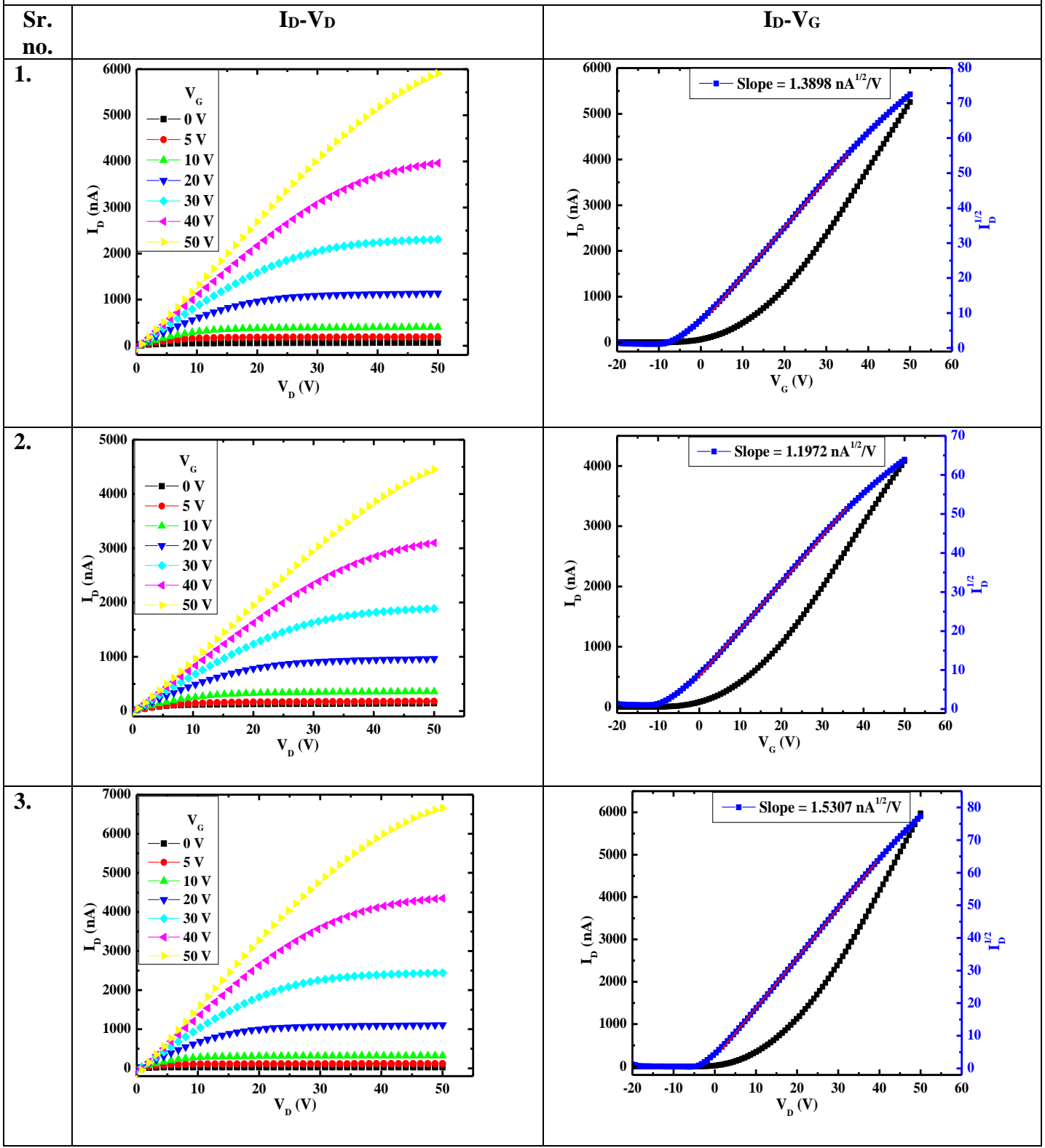

| 4. | 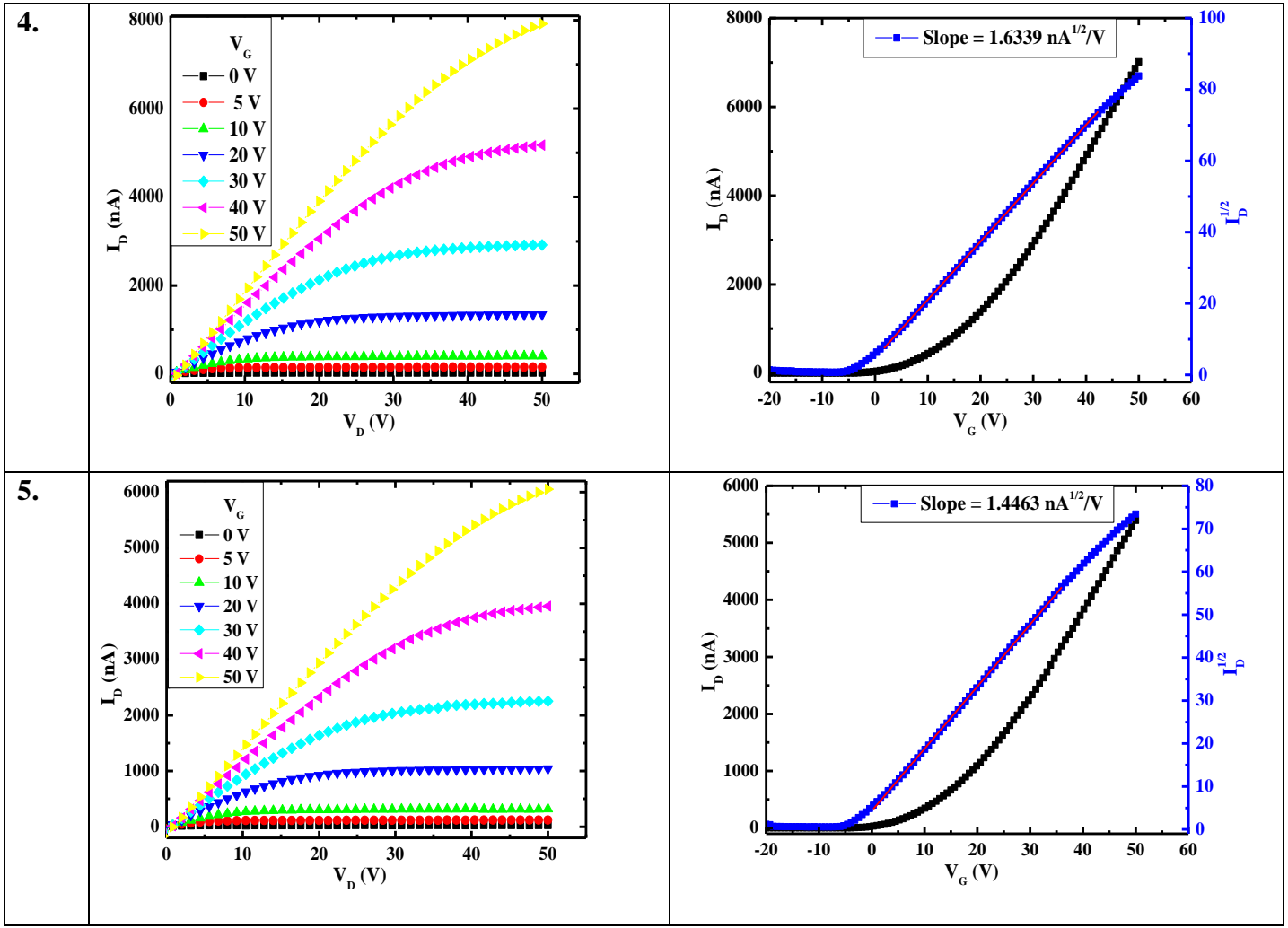 | |

| 5. | | |

| 5) 0.5% PDI | | |
|---|---|---|
| Sr. no. | $I_D$-$V_D$ | $I_D$-$V_G$ |
| 1. | 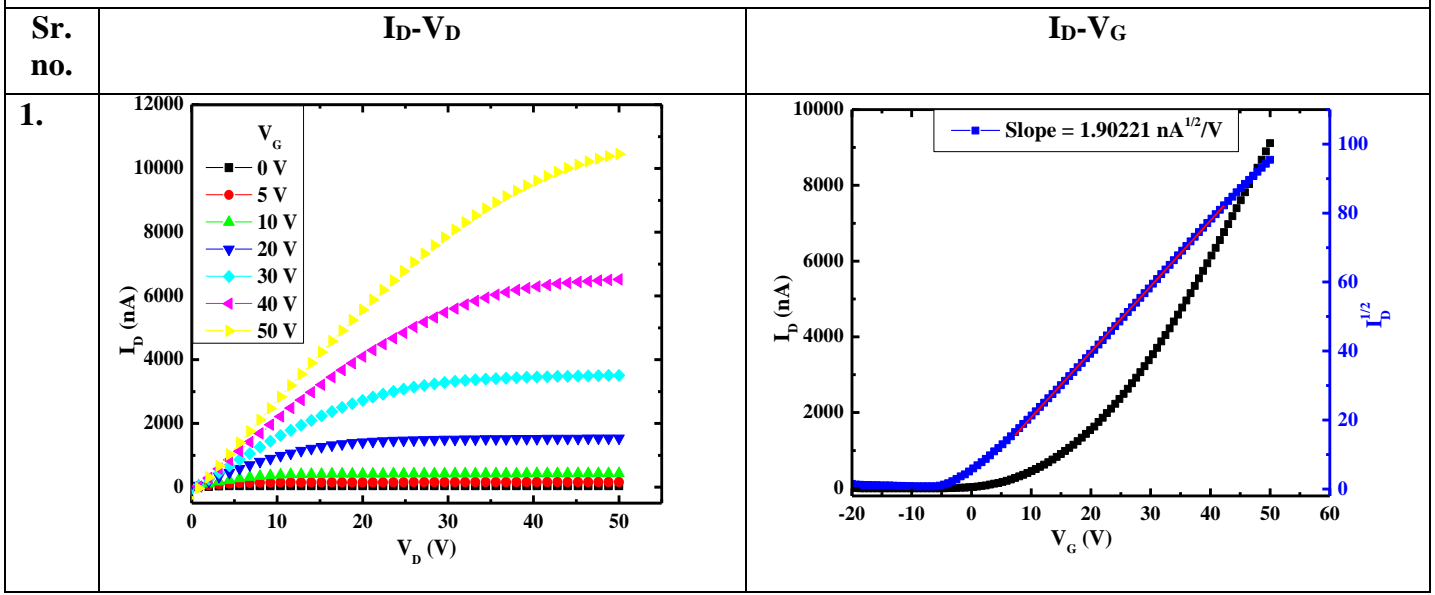 | |

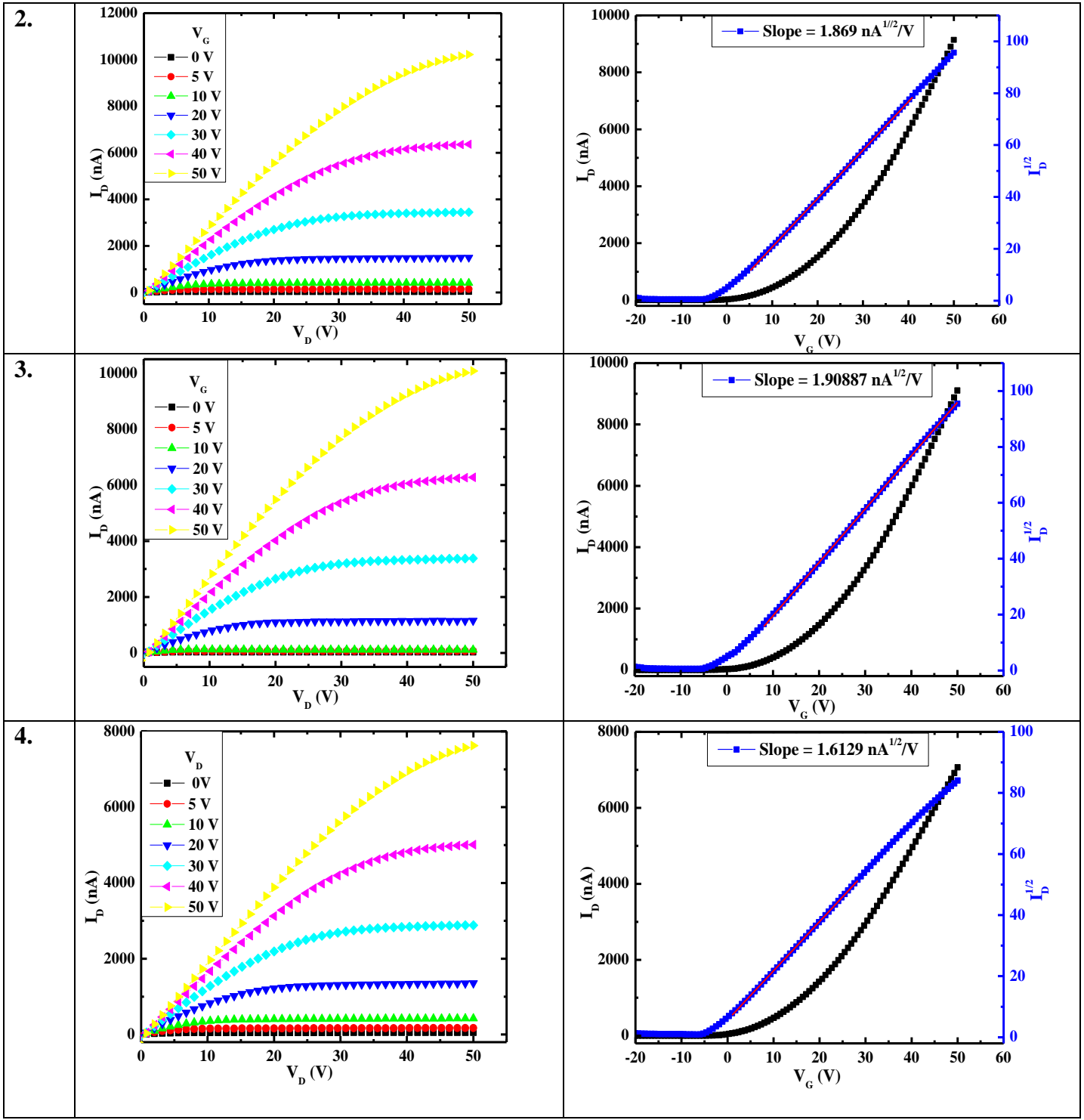

| 5. | 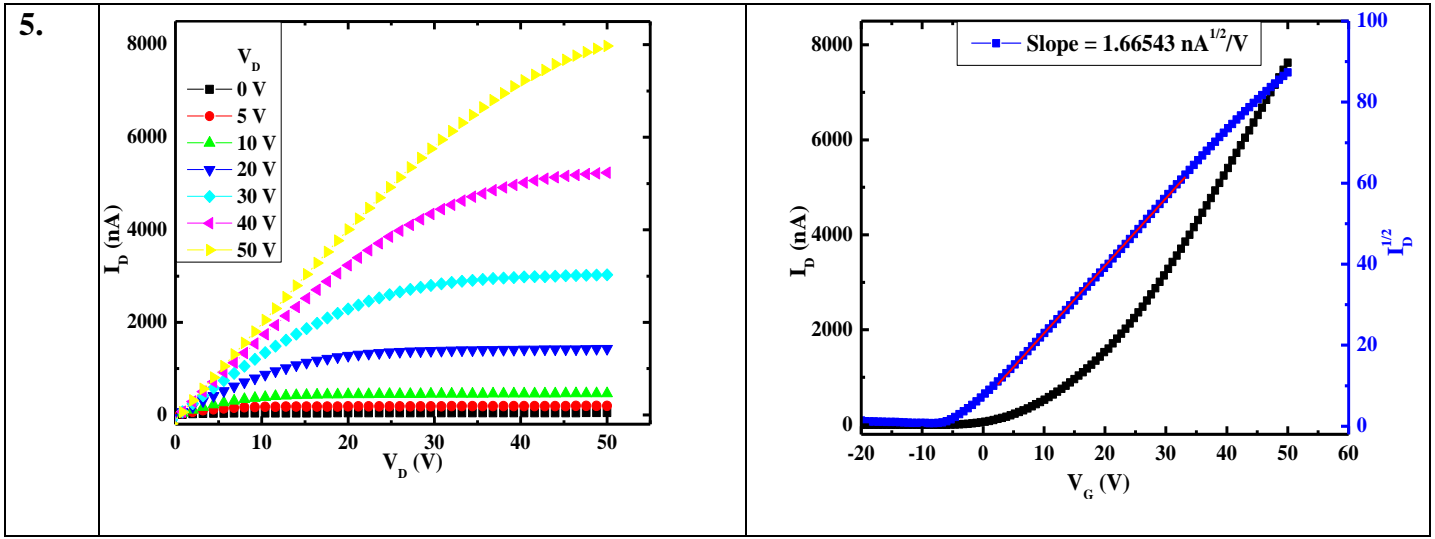 | |

| | 6) 0.8% PDI | |
|---|---|---|
| Sr. no. | $I_D$-$V_D$ | $I_D$-$V_G$ |
| 1. | 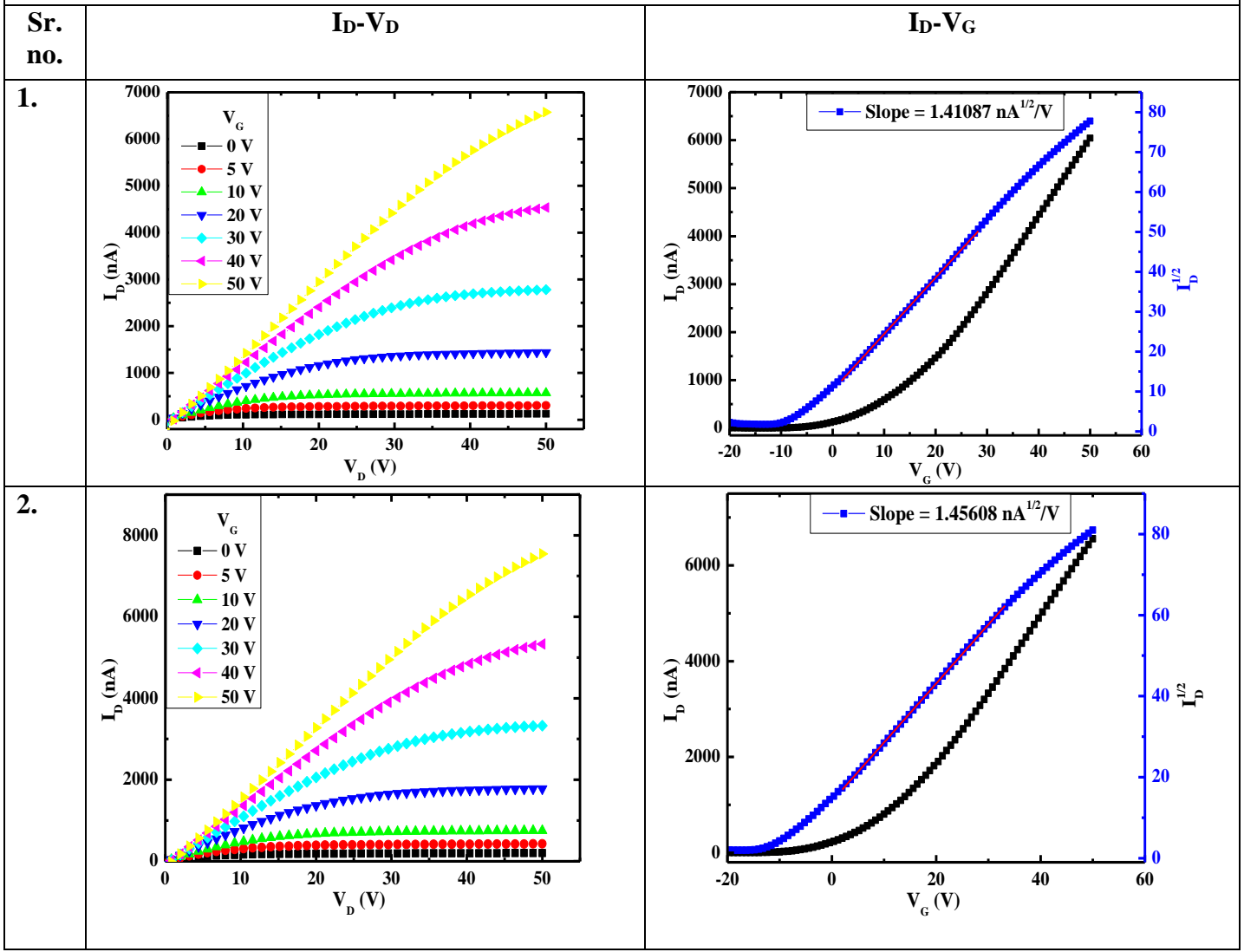 | |
| 2. | | |

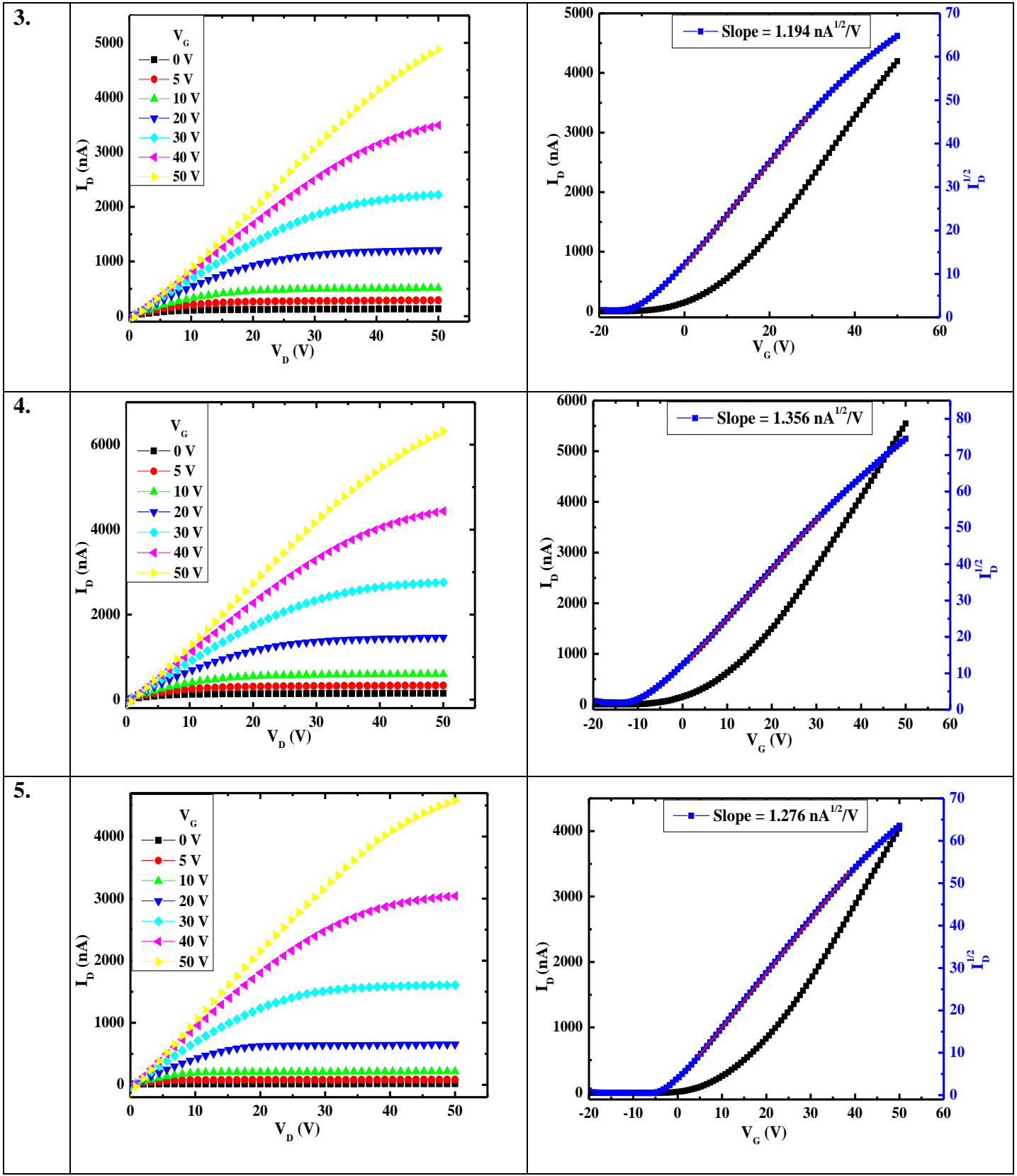

| | 7) 1% PDI | |
|---|---|---|
| Sr. no. | $I_D$-$V_D$ | $I_D$-$V_G$ |
| 1. | | |
| 2. | | |
| 3. | | |

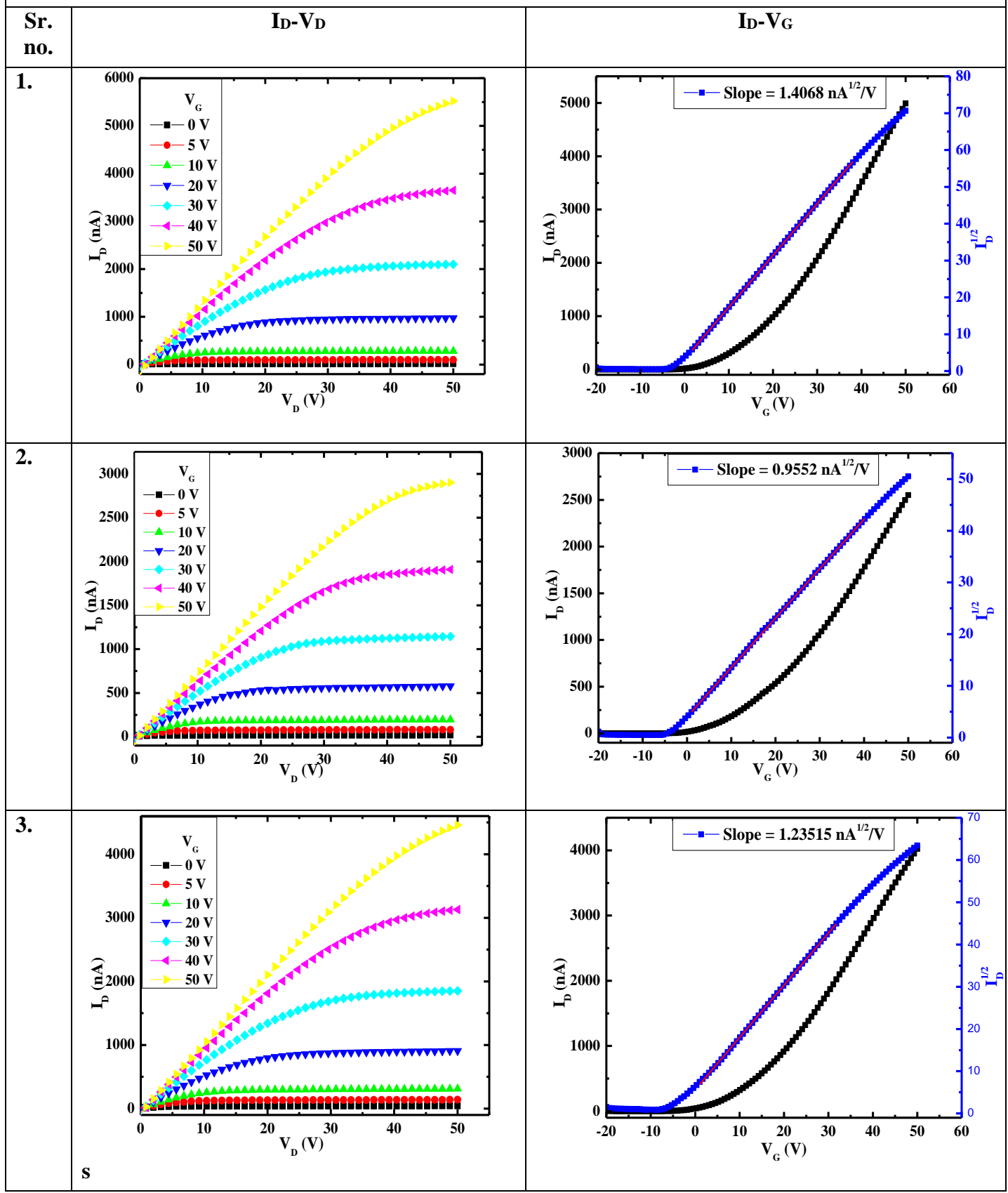

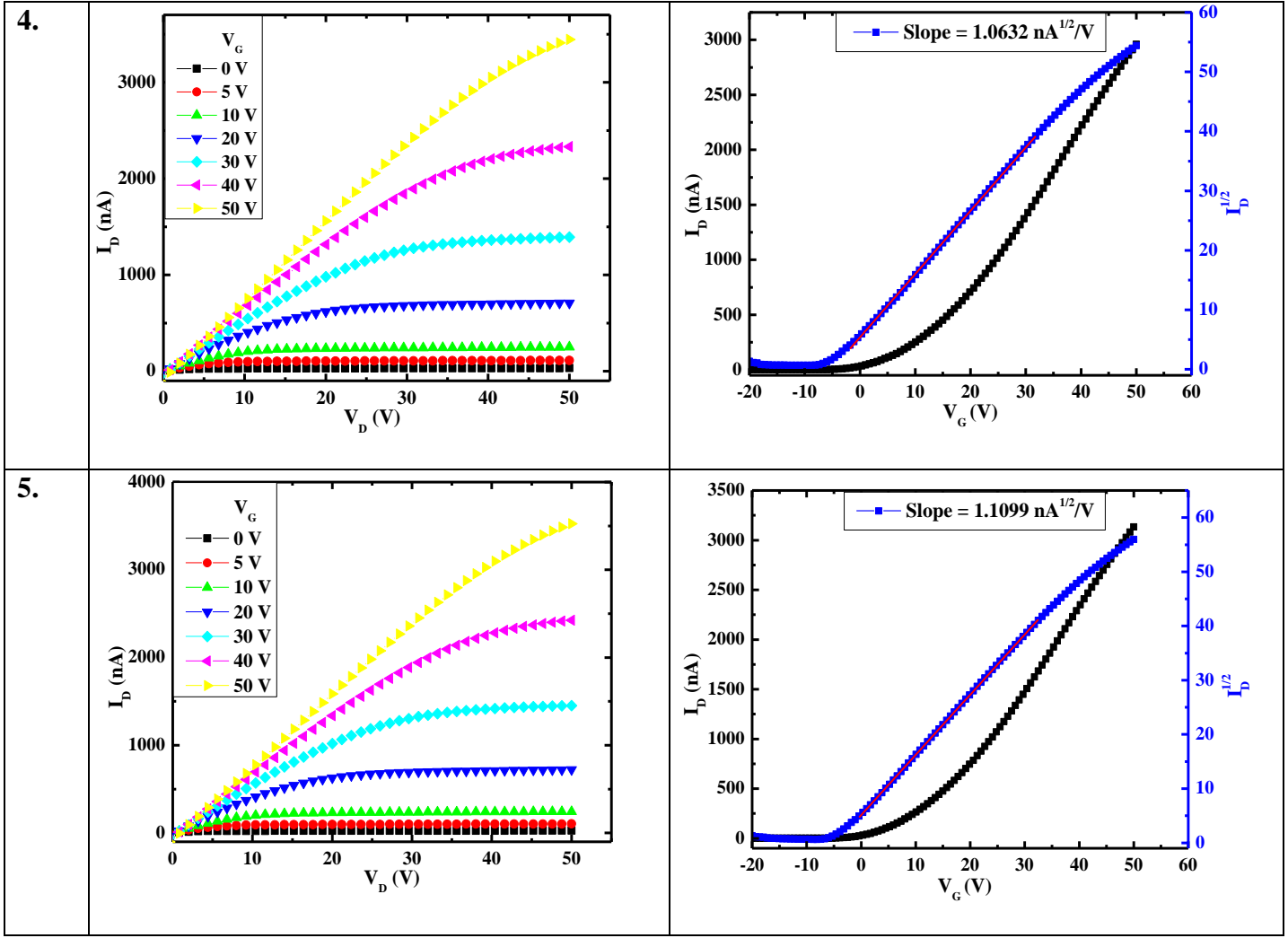

## Section-8

**Averaged data for λ/2 and λ-cavity:** Slope and mobility were extracted from electrical measurements. The thickness of PDI film in each sample is estimated by TMM fitting of each sample's reflectance spectra. Empty cavity mode positions were also calculated from TMM. Then average and the standard deviation are computed for both λ/2 and λ-cavities.

I. **For λ/2-Cavity**

**T1-1**

| | 1) 0.05% PDI | | | |
|---|---|---|---|---|
| Sample number | Empty mode position (nm) (from TMM) | Thickness (nm) (from TMM) | Slope (nA$^{1/2}$ V$^{-1}$) | Mobility (x 10$^{-3}$ cm$^2$V$^{-1}$s$^{-1}$) |
| 1 | 575 | 8 | 3.32 | 5.7408 |
| 2 | 590 | 10 | 3.37 | 5.91501 |
| 3 | 590 | 10 | 4.187 | 9.13065 |
| 4 | 575 | 8 | 2.806 | 4.10083 |
| 5 | 575 | 8 | 2.9745 | 4.60812 |
| 6 | 575 | 8 | 3.5475 | 6.55452 |
| 7 | 575 | 8 | 3.6468 | 6.9266 |
| 8 | 575 | 8 | 4.574 | 10.89653 |
| 9 | 575 | 8 | 3.9626 | 8.17818 |
| 10 | 575 | 8 | 3.857 | 7.7481 |
| 11 | 575 | 8 | 4.2875 | 9.57424 |
| 12 | 575 | 8 | 3.4786 | 6.30239 |
| 13 | 570 | 7 | 2.572 | 3.44539 |
| 14 | 570 | 7 | 3.0819 | 4.9469 |
| 15 | 570 | 7 | 3.4346 | 6.14396 |
| 16 | 570 | 7 | 3.25 | 5.50127 |
| 17 | 570 | 7 | 3.45 | 6.19918 |
| 18 | 570 | 7 | 3.719 | 7.20358 |
| 19 | 570 | 7 | 2.632 | 3.60801 |
| 20 | 570 | 7 | 2.88 | 4.31997 |
| 21 | 585 | 9 | 2.49 | 3.2292 |
| 22 | 585 | 9 | 2.989 | 4.65316 |
| 23 | 585 | 9 | 3.0936 | 4.98453 |
| 24 | 585 | 9 | 2.959 | 4.56022 |
| 25 | 585 | 9 | 3.1797 | 5.26585 |
| 26 | 585 | 9 | 2.994 | 4.66874 |
| 27 | 585 | 9 | 3.383 | 5.96074 |
| 28 | 585 | 9 | 3.204 | 5.34664 |
| 29 | 585 | 9 | 2.583 | 3.47492 |
| 30 | 585 | 9 | 2.875 | 4.30499 |
| 31 | 585 | 9 | 3.5499 | 6.56339 |
| 32 | 585 | 9 | 2.45 | 3.12628 |
| 33 | 585 | 9 | 4.25 | 9.40749 |
| 34 | 585 | 9 | 3.169 | 5.23047 |
| 35 | 585 | 9 | 3.653 | 6.95017 |
| 36 | 585 | 9 | 4.031 | 8.46295 |
| 37 | 585 | 9 | 3.0247 | 4.76497 |
| 38 | 585 | 9 | 3.5679 | 6.63012 |
| **Average** | **579.47368** | **8.368** | **3.32834** | **5.91129** |
| **SD** | **6.66782** | **0.87121** | **0.52146** | **1.87168** |

**T1-2**

| | 2) 0.1% PDI | | | |
|---|---|---|---|---|
| Sample number | Empty mode position (nm) (from TMM) | Thickness (nm) (from TMM) | Slope (nA$^{1/2}$ V$^{-1}$) | Mobility (x 10$^{-3}$ cm$^2$V$^{-1}$s$^{-1}$) |
| 1 | 685 | 20 | 2.842 | 4.20673 |
| 2 | 685 | 20 | 3.198 | 5.32663 |
| 3 | 660 | 18 | 1.798 | 1.68374 |
| 4 | 685 | 20 | 2.72 | 3.85331 |
| 5 | 685 | 20 | 2.6449 | 3.64346 |
| 6 | 645 | 15 | 1.9801 | 2.04207 |
| 7 | 645 | 15 | 1.9813 | 2.04454 |
| 8 | 645 | 15 | 2.317 | 2.79607 |
| 9 | 645 | 15 | 1.8967 | 1.87367 |
| 10 | 675 | 19 | 2.7557 | 3.95512 |
| 11 | 675 | 19 | 3.1705 | 5.23542 |
| 12 | 645 | 15 | 1.9422 | 1.96464 |
| 13 | 645 | 15 | 1.453 | 1.09958 |
| 14 | 645 | 15 | 1.902 | 1.88416 |
| 15 | 645 | 15 | 2.3629 | 2.90795 |
| **Average** | **660.667** | **17.067** | **2.33095** | **2.96781** |
| **SD** | **17.78264** | **2.26471** | **0.51467** | **1.28364** |

**T1-3**

| | 3) 0.2% PDI | | | |
|---|---|---|---|---|
| Sample number | Empty mode position (nm) (from TMM) | Thickness (nm) (from TMM) | Slope (nA$^{1/2}$ V$^{-1}$) | Mobility (x 10$^{-3}$ cm$^2$V$^{-1}$s$^{-1}$) |
| 1 | 780 | 32 | 2.318 | 2.79848 |
| 2 | 780 | 32 | 2.629 | 3.59979 |
| 3 | 740 | 27 | 1.871 | 1.82324 |
| 4 | 740 | 27 | 1.9759 | 2.03341 |
| 5 | 710 | 23 | 1.2067 | 0.75839 |
| 6 | 710 | 23 | 1.401 | 1.02229 |
| 7 | 710 | 23 | 1.258 | 0.82425 |
| 8 | 710 | 23 | 1.223 | 0.77902 |
| 9 | 735 | 26 | 1.54 | 1.2352 |
| 10 | 735 | 26 | 1.64332 | 1.4065 |
| **Average** | **735** | **26.2** | **1.70659** | **1.62806** |
| **SD** | **25.69047** | **3.31059** | **0.46199** | **0.90298** |

**T1-4**

| | 4) 0.3% PDI | | | |
|---|---|---|---|---|
| Sample number | Empty mode position (nm) (from TMM) | Thickness (nm) (from TMM) | Slope ($nA^{1/2} V^{-1}$) | Mobility (x $10^{-3}$ $cm^2V^{-1}s^{-1}$) |
| 1 | 810 | 35 | 2.23 | 2.59004 |
| 2 | 810 | 35 | 1.87 | 1.82129 |
| 3 | 750 | 28 | 1.0622 | 0.58764 |
| 4 | 750 | 28 | 1.37 | 0.97755 |
| 5 | 750 | 28 | 1.06 | 0.5852 |
| 6 | 750 | 28 | 1.22 | 0.7752 |
| 7 | 760 | 30 | 1.7208 | 1.54226 |
| 8 | 760 | 30 | 1.39528 | 1.01396 |
| 9 | 780 | 32 | 1.50989 | 1.18737 |
| 10 | 750 | 28 | 1.57039 | 1.28443 |
| **Average** | **767** | **30.2** | **1.50086** | **1.23649** |
| **SD** | **23.25941** | **2.71293** | **0.34859** | **0.5871** |

**T1-5**

| | 5) 0.5% PDI | | | |
|---|---|---|---|---|
| Sample number | Empty mode position (nm) (from TMM) | Thickness (nm) (from TMM) | Slope ($nA^{1/2} V^{-1}$) | Mobility (x $10^{-3}$ $cm^2V^{-1}s^{-1}$) |
| 1 | 955 | 60 | 1.9258 | 1.93161 |
| 2 | 955 | 60 | 2.0907 | 2.27656 |
| 3 | 955 | 60 | 1.4037 | 1.02623 |
| 4 | 955 | 60 | 2.0824 | 2.25852 |
| 5 | 900 | 45 | 1.887 | 1.85456 |
| 6 | 900 | 45 | 1.784 | 1.65762 |
| 7 | 900 | 45 | 1.424 | 1.05613 |
| 8 | 900 | 45 | 1.5566 | 1.26197 |
| 9 | 930 | 50 | 2.0737 | 2.23969 |
| 10 | 930 | 50 | 1.7312 | 1.56096 |
| 11 | 930 | 50 | 2.4227 | 3.057 |
| 12 | 930 | 50 | 1.6732 | 1.45811 |
| 13 | 930 | 50 | 1.7204 | 1.54154 |
| 14 | 930 | 50 | 1.0614 | 0.58675 |
| 15 | 930 | 50 | 1.51 | 1.18754 |
| 16 | 930 | 50 | 1.6984 | 1.50237 |
| 17 | 930 | 50 | 1.5833 | 1.30564 |
| 18 | 930 | 50 | 1.6379 | 1.39724 |
| 19 | 930 | 50 | 1.2345 | 0.79374 |
| 20 | 930 | 50 | 1.1533 | 0.69276 |
| 21 | 930 | 50 | 1.717 | 1.53545 |
| 22 | 930 | 50 | 1.494 | 1.16251 |
| 23 | 930 | 50 | 1.964 | 2.009 |
| 24 | 930 | 50 | 1.9246 | 1.9292 |
| 25 | 920 | 48 | 1.5259 | 1.21269 |
| 26 | 920 | 48 | 1.7094 | 1.52189 |
| 27 | 920 | 48 | 1.53 | 1.21921 |
| 28 | 920 | 48 | 2.023 | 2.13151 |
| 29 | 920 | 48 | 1.623 | 1.37193 |
| 30 | 920 | 48 | 1.86 | 1.80186 |
| 31 | 920 | 48 | 1.6978 | 1.50131 |
| 32 | 920 | 48 | 1.8196 | 1.72444 |
| 33 | 920 | 48 | 1.64 | 1.40082 |
| 34 | 930 | 50 | 1.8044 | 1.69575 |
| 35 | 930 | 50 | 1.63 | 1.38379 |
| 36 | 930 | 50 | 1.589 | 1.31505 |
| **Average** | **926.79** | **50.056** | **1.70016** | **1.54342** |
| **SD** | **4.6702** | **3.85821** | **0.26985** | **0.48326** |

## II. For λ-Cavity

**T2-1**

| 1) 0.05% PDI ||||  |
|---|---|---|---|---|
| Sample number | Empty mode position (nm) (from TMM) | Thickness (nm) (from TMM) | Slope ($nA^{1/2} V^{-1}$) | Mobility (x $10^{-3} cm^2 V^{-1} s^{-1}$) |
| 1 | 485 | 8 | 1.78 | 4.22348 |
| 2 | 485 | 8 | 1.92 | 4.91397 |
| 3 | 485 | 8 | 1.88 | 4.71136 |
| 4 | 485 | 8 | 1.85 | 4.56219 |
| 5 | 485 | 8 | 2.079 | 5.76155 |
| 6 | 485 | 8 | 2.13 | 6.04769 |
| 7 | 485 | 8 | 2.1638 | 6.24115 |
| 8 | 478 | 7 | 2.2912 | 6.99771 |
| 9 | 478 | 7 | 2.4477 | 7.98632 |
| 10 | 478 | 7 | 2.50 | 8.33125 |
| 11 | 478 | 7 | 2.0907 | 5.82658 |
| 12 | 478 | 7 | 2.1819 | 6.346 |
| **Average** | **482** | **7.58** | **2.10952** | **5.99577** |
| **SD** | **3.45105** | **0.49301** | **0.21876** | **1.24853** |

**T2-2**

| 2) 0.1% PDI | | | | |
|---|---|---|---|---|
| Sample number | Empty mode position (nm) (from TMM) | Thickness (nm) (from TMM) | Slope ($nA^{1/2} V^{-1}$) | Mobility (x $10^{-3} cm^2V^{-1}s^{-1}$) |
| 1 | 510 | 16 | 1.149 | 1.75983 |
| 2 | 510 | 16 | 1.282 | 2.19082 |
| 3 | 510 | 16 | 1.226 | 2.0036 |
| 4 | 510 | 16 | 1.2603 | 2.11728 |
| 5 | 492 | 10 | 1.75 | 4.08231 |
| 6 | 492 | 10 | 1.748 | 4.07299 |
| 7 | 492 | 10 | 1.589 | 3.36572 |
| 8 | 492 | 10 | 1.83 | 4.46408 |
| 9 | 492 | 10 | 1.635 | 3.56341 |
| 10 | 492 | 10 | 1.526 | 3.10413 |
| 11 | 492 | 10 | 1.515 | 3.05953 |
| 12 | 492 | 10 | 1.559 | 3.23983 |
| 13 | 492 | 10 | 1.476 | 2.90404 |
| 14 | 505 | 15 | 1.32 | 2.32262 |
| 15 | 505 | 15 | 1.3048 | 2.26944 |
| 16 | 505 | 15 | 1.3326 | 2.36717 |
| 17 | 505 | 15 | 1.3489 | 2.42544 |
| 18 | 492 | 10 | 1.3736 | 2.51507 |
| 19 | 492 | 10 | 1.5127 | 3.05025 |
| **Average** | **498.52** | **12.3** | **1.45989** | **2.88829** |
| **SD** | **7.82283** | **2.73482** | **0.18837** | **0.74999** |

**T2-3**

| 3) 0.2% PDI | | | | |
|---|---|---|---|---|
| Sample number | Empty mode position (nm) (from TMM) | Thickness (nm) (from TMM) | Slope ($nA^{1/2} V^{-1}$) | Mobility (x $10^{-3} cm^2V^{-1}s^{-1}$) |
| 1 | 555 | 33 | 0.79558 | 0.84372 |
| 2 | 555 | 33 | 0.818 | 0.89194 |
| 3 | 535 | 25 | 0.906 | 1.09417 |
| 4 | 535 | 25 | 1.07 | 1.52615 |
| 5 | 535 | 25 | 0.7288 | 0.70802 |
| 6 | 535 | 25 | 1.007 | 1.35173 |
| 7 | 535 | 25 | 0.9963 | 1.32315 |
| 8 | 535 | 25 | 0.9922 | 1.31229 |
| 9 | 535 | 25 | 0.8694 | 1.00756 |
| 10 | 535 | 25 | 0.8407 | 0.94213 |
| **Average** | **539** | **26.6** | **0.9024** | **1.10009** |
| **SD** | **8** | **3.2** | **0.10464** | **0.25218** |

**T2-4**

| | 4) | | 0.3% PDI | |
|---|---|---|---|---|
| Sample number | Empty mode position (nm) (from TMM) | Thickness (nm) (from TMM) | Slope ($nA^{1/2} V^{-1}$) | Mobility (x $10^{-3} cm^2V^{-1}s^{-1}$) |
| 1 | 586 | 48 | 1.389 | 2.57178 |
| 2 | 586 | 48 | 1.197 | 1.90993 |
| 3 | 580 | 45 | 1.53 | 3.12042 |
| 4 | 580 | 45 | 1.6339 | 3.55862 |
| 5 | 580 | 45 | 1.446 | 2.78719 |
| 6 | 580 | 45 | 1.5049 | 3.01888 |
| **Average** | **582** | **46** | **1.45013** | **2.8278** |
| **SD** | **2.82843** | **1.41421** | **0.136** | **0.51087** |

**T2-5**

| | 5) | | 0.5% PDI | |
|---|---|---|---|---|
| Sample number | Empty mode position (nm) (from TMM) | Thickness (nm) (from TMM) | Slope ($nA^{1/2} V^{-1}$) | Mobility (x $10^{-3} cm^2V^{-1}s^{-1}$) |
| 1 | 608 | 60 | 0.918 | 1.12335 |
| 2 | 613 | 63 | 1.63 | 3.54165 |
| 3 | 613 | 63 | 1.78 | 4.22348 |
| 4 | 613 | 63 | 1.77 | 4.17616 |
| 5 | 598 | 55 | 1.263 | 2.12636 |
| 6 | 598 | 55 | 1.552 | 3.2108 |
| 7 | 598 | 55 | 1.116 | 1.66019 |
| 8 | 600 | 57 | 1.7158 | 3.92431 |
| 9 | 608 | 60 | 1.9179 | 4.90323 |
| 10 | 600 | 57 | 1.67 | 3.7176 |
| 11 | 587 | 50 | 1.8547 | 4.5854 |
| 12 | 600 | 56 | 1.930 | 4.96529 |
| 13 | 600 | 56 | 1.902 | 4.82227 |
| 14 | 600 | 56 | 1.869 | 4.65638 |
| 15 | 600 | 56 | 1.909 | 4.85783 |
| 16 | 598 | 55 | 1.613 | 3.46816 |
| 17 | 598 | 55 | 1.665 | 3.69538 |
| **Average** | **601.88** | **57.18** | **1.60914** | **3.74458** |
| **SD** | **6.70279** | **3.41683** | **0.41537** | **0.12353** |

**T2-6**

| | 6)    0.8% PDI | | | |
|---|---|---|---|---|
| Sample number | Empty mode position (nm) (from TMM) | Thickness (nm) (from TMM) | Slope (nA$^{1/2}$ V$^{-1}$) | Mobility (x 10$^{-3}$ cm$^2$V$^{-1}$s$^{-1}$) |
| 1 | 612 | 61 | 1.41 | 2.65014 |
| 2 | 612 | 60 | 1.456 | 2.82587 |
| 3 | 608 | 60 | 1.194 | 1.90037 |
| 4 | 612 | 61 | 1.356 | 2.45104 |
| 5 | 613 | 63 | 1.108 | 1.63648 |
| 6 | 613 | 63 | 1.276 | 2.17036 |
| 7 | 615 | 62 | 1.3739 | 2.51617 |
| **Average** | **612.14** | **61.428** | **1.31056** | **2.3072** |
| **SD** | **1.95876** | **1.17803** | **0.11522** | **0.39372** |

**T2-7**

| | 7)    1% PDI | | | |
|---|---|---|---|---|
| Sample number | Empty mode position (nm) (from TMM) | Thickness (nm) (from TMM) | Slope (nA$^{1/2}$ V$^{-1}$) | Mobility (x 10$^{-3}$ cm$^2$V$^{-1}$s$^{-1}$) |
| 1 | 625 | 67 | 1.4068 | 2.63812 |
| 2 | 621 | 65 | 0.955 | 1.21573 |
| 3 | 620 | 64 | 1.235 | 2.03312 |
| 4 | 620 | 64 | 1.060 | 1.49776 |
| 5 | 621 | 65 | 1.109 | 1.63943 |
| **Average** | **621.4** | **65** | **1.15316** | **1.80483** |
| **SD** | **1.85472** | **1.09545** | **0.5552** | **0.49299** |

## Supporting references: